%% file: main.tex
\documentclass[sigconf]{acmart}

\AtBeginDocument{%
  }
    
\input{preamble/copyright.tex}
\input{preamble/packages.tex}

\input{preamble/definitions.tex}

\begin{document}

\title{Reproducing NevIR: Negation in Neural Information Retrieval}

\input{preamble/authors}
\input{sections/00_abstract}
\input{sections/keywords}

\maketitle

\input{sections/01_introduction}
\input{sections/02_related_work}
\input{sections/03_nevir_overview}
\input{sections/04_methodology}
\input{sections/05_experimental_results}

\input{sections/06_discussion_conclusions}
\input{sections/07_acknowledgments}

\newpage
\bibliographystyle{ACM-Reference-Format}
\balance
\bibliography{references}

% \appendix

\end{document}

%% file: preamble/copyright.tex
\copyrightyear{2025}
\acmYear{2025}
\setcopyright{cc}
\setcctype{by}
\acmConference[SIGIR '25]{Proceedings of the 48th International ACM SIGIR
Conference on Research and Development in Information Retrieval}{July 13--18,
2025}{Padua, Italy}
\acmBooktitle{Proceedings of the 48th International ACM SIGIR Conference on
Research and Development in Information Retrieval (SIGIR '25), July 13--18,
2025, Padua, Italy}\acmDOI{10.1145/3726302.3730294}
\acmISBN{979-8-4007-1592-1/2025/07}

%% file: preamble/packages.tex
\usepackage[inline]{enumitem}
\usepackage{todonotes}
\newlist{inlinelist}{enumerate*}{1}
\setlist*[inlinelist,1]{label=\roman*),itemjoin={{, }},itemjoin*={{, and }}}
\usepackage{multirow}

\usepackage[T1]{fontenc}
\usepackage[htt]{hyphenat}

%% file: preamble/definitions.tex
\usepackage{xcolor}
\definecolor{alsopink}{RGB}{255, 0, 255}

\newcommand{\header}[1]{\vspace{2mm}\noindent\textbf{#1}}

%% file: preamble/authors.tex

\author{Coen van den Elsen}
\authornote{Equal contributions.}
\affiliation{%
  \institution{University of Amsterdam}
  \city{Amsterdam}
  \country{The Netherlands}
}
\email{coen.van.den.elsen@student.uva.nl}

\author{Francien Barkhof}
\authornotemark[1]
\affiliation{%
  \institution{University of Amsterdam}
  \city{Amsterdam}
  \country{The Netherlands}
}
\email{francien.barkhof@student.uva.nl}

\author{Thijmen Nijdam}
\authornotemark[1]
\affiliation{%
  \institution{University of Amsterdam}
  \city{Amsterdam}
  \country{The Netherlands}
}
\email{thijmen.nijdam@student.uva.nl}

\author{Simon Lupart}
\affiliation{%
  \institution{University of Amsterdam}
  \city{Amsterdam}
  \country{The Netherlands}
}
\email{s.c.lupart@uva.nl}

\author{Mohammad Aliannejadi}
\affiliation{%
  \institution{University of Amsterdam}
  \city{Amsterdam}
  \country{The Netherlands}
}
\email{m.aliannejadi@uva.nl}

% \renewcommand{\shortauthors}{Barkhof et al.}
\renewcommand{\shortauthors}{Coen van den Elsen, Francien Barkhof, Thijmen Nijdam, Simon Lupart, and Mohammad Aliannejadi}

%% file: sections/00_abstract.tex

\begin{abstract}
Negation is a fundamental aspect of human communication, yet it remains a challenge for Language Models (LMs) in Information Retrieval (IR). Despite the heavy reliance of modern neural IR systems on LMs, little attention has been given to their handling of negation. In this study, we reproduce and extend the findings of NevIR, a benchmark study that revealed most IR models perform at or below the level of random ranking when dealing with negation. We replicate NevIR’s original experiments and evaluate newly developed state-of-the-art IR models. Our findings show that a recently emerging category---listwise Large Language Model (LLM) re-rankers—outperforms other models but still underperforms human performance. Additionally, we leverage ExcluIR, a benchmark dataset designed for exclusionary queries with extensive negation, to assess the generalisability of negation understanding. Our findings suggest that fine-tuning on one dataset does not reliably improve performance on the other, indicating notable differences in their data distributions. Furthermore,
we observe that only cross-encoders and listwise LLM re-rankers achieve reasonable performance across both negation tasks.
\end{abstract}

%% file: sections/keywords.tex
\begin{CCSXML}
<ccs2012>
   <concept>
       <concept_id>10002951.10003317</concept_id>
       <concept_desc>Information systems~Information retrieval</concept_desc>
       <concept_significance>500</concept_significance>
       </concept>
   <concept>
       <concept_id>10002951.10003317.10003359</concept_id>
       <concept_desc>Information systems~Evaluation of retrieval results</concept_desc>
       <concept_significance>500</concept_significance>
       </concept>
   <concept>
       <concept_id>10002951.10003317.10003359.10003360</concept_id>
       <concept_desc>Information systems~Test collections</concept_desc>
       <concept_significance>500</concept_significance>
       </concept>
 </ccs2012>
\end{CCSXML}

\ccsdesc[500]{Information systems~Information retrieval}
\ccsdesc[500]{Information systems~Evaluation of retrieval results}
\ccsdesc[500]{Information systems~Test collections}

\keywords{Negation, Information Retrieval, Benchmark}

%% file: sections/01_introduction.tex
\section{Introduction}
\begin{figure}
    \centering
    \includegraphics[width=\linewidth]{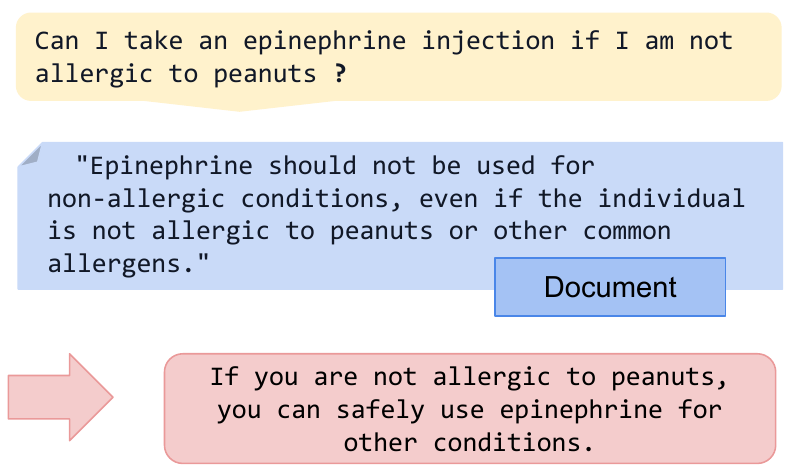}
    \caption{This example illustrates how misinterpreting a negation in a medical context could cause someone to misuse a critical emergency medication like epinephrine}
    \label{fig:intro_example}
\end{figure}

Negation plays a crucial role in human communication, enabling the expression of contradictions, exclusions, and oppositions. Despite its common use in natural language, effectively handling negation remains a significant challenge for Language Models (LMs)~\cite{condaqa}. This issue has notable implications for Information Retrieval (IR) tasks, as different and complex forms of negation can appear in both user queries and documents~\cite{condaqa}. This misinterpretation of negation is undesirable and can be harmful, particularly in high-risk retrieval tasks such as healthcare or legal settings, where it can lead to serious errors~\cite{weller2024nevirnegationneuralinformation}. For example, as shown in \autoref{fig:intro_example}, misinterpreting negation in the medical question-answering domain can have severe and even catastrophic consequences. As can be seen in \autoref{fig:intro_example}, missing the negation on the ``non-allergic'' cases would lead to a catastrophic recommendation generated by the system that prescribes epinephrine to the user with no peanuts allergy.

Prior work on negation in IR, primarily from the pre-neural era, relied on rule-based and early statistical methods~\cite{chapman2001simple, harkema2009context, goldin2003learning, uzuner20112010}, focusing on explicit lexical cues. However, these methods largely overlooked more nuanced or implicit forms of negation. Despite its importance, there has been little research on how modern neural IR models handle negation. To address this gap,~\citet{weller2024nevirnegationneuralinformation} propose a novel benchmark dataset —NevIR— to specifically evaluate how more recent neural IR systems handle negation. The NevIR dataset consists of pairs of queries, each semantically aligned with exactly one of two contrastive documents. These document pairs differ only in the presence or absence of negation terms (e.g., ``The box is locked.'' vs.\ ``The box is unlocked.''). The authors tested neural IR models by assessing their ability to correctly rank the appropriate document above its negated counterpart. The results showed that most state-of-the-art (SotA) IR models struggle with negation, often performing worse than random ranking. Given these limitations identified by the NevIR benchmark, an important question arises: do these findings generalise to newer IR model families, particularly LLM-based IR methods? Recently, LLMs have demonstrated strong effectiveness in IR, making them promising candidates for retrieval tasks~\cite{repllama_rankllama, rankGPT, qin2023large, ma2023zeroshotlistwisedocumentreranking}, but their ability to process negation remains uncertain. 

In this work, we therefore not only reproduce the experiments of the NevIR paper~\cite{weller2024nevirnegationneuralinformation}, but also extend its evaluation to LLM-based retrieval methods.  In doing so, we aim to assess whether recent advances in IR have improved negation handling or whether it remains a persistent challenge. 

Furthermore,  \citet{weller2024nevirnegationneuralinformation} propose a preliminary solution for handling negation by fine-tuning IR models on the NevIR dataset. While this improves performance significantly, it remains unclear whether models truly learn negation patterns or overfit the data distribution of the NevIR dataset. This is particularly interesting given that the dataset is created synthetically and could potentially lack negation and language diversity. To address this, we build on the recent work of \citet{zhang2024excluir}, who introduced ExcluIR—a benchmark focused specifically on exclusionary queries. For example, the query ``What other flowers besides tulips grow in the Netherlands?'' expresses exclusion by implicitly negating one category. By fine-tuning and evaluating models on both NevIR and ExcluIR, we explore the generalisation of negation handling across datasets and assess the effectiveness of NevIR in helping models understand negation in IR.

Through this study, we seek to extend the current work on negation in IR, contributing to a clearer understanding of how different datasets and models affect negation understanding. Additionally, we aim to extend the initial NevIR experiments by exploring the potential trade-offs between improving negation understanding and preserving ranking performance. We address the following research questions:
\begin{enumerate}[label=\textbf{RQ\arabic*}]
    \item How do the negation understanding results of the NevIR dataset generalise to the most recent retrieval architectures, such as listwise LLM re-rankers?
    \item How do listwise LLM re-rankers perform when fine-tuned on the negation data of NevIR?
    \item How does fine-tuning on other negation datasets, such as ExcluIR, improve generalisability on NevIR?
    \item How can performance on a negation dataset be improved while maintaining strong ranking capabilities?
\end{enumerate}

In response to these questions, we present the following findings:

\begin{itemize}[leftmargin=*]
    % RQ1
    \item We observe a positive trend in handling negation on newly developed SotA IR models. Notably, a newly emerged model category, listwise LLM re-rankers, achieved 20\% gains compared to the previous approaches. We also observe that the recent advances in cross-encoders and bi-encoders with models such as \url{GritLM} and \url{jina-reranker-v2-base-multilingual} show an improvement in negation understanding.
    \item When fine-tuning, we find that, despite its superior base performance, listwise LLM re-rankers have similar learning capabilities for the negation task as other model categories.
    \item Fine-tuning experiments on NevIR and ExcluIR reveal that only the model from the cross-encoder category can generalise between the two negation datasets. 
    \item Our analysis of training IR models on negation data shows that a trade-off selection criterion can help mitigate overfitting.
\end{itemize}

Our code, which ensures full reproducibility of our experiments, is publicly available on GitHub\footnote{\url{https://github.com/thijmennijdam/NevIR-reproducibility}}.

%% file: sections/02_related_work.tex
\section{Related Work}

\header{Negation in NLP.}
Negation is a complex semantic phenomenon in natural language that affects the performance of various Natural Language Processing (NLP) applications \cite{sineva2021negation,morante2021recent}. It ``transforms an expression into another expression whose meaning is in some way opposed to the original'' as mentioned by \citet{morante2021recent}. It occurs frequently, with the proportion of sentences with negation in English corpora ranging between 9\% and 32\% \cite{sineva2021negation}. Negation is a challenging problem in machine translation, natural language inference, and sentiment analysis \cite{garcia2023not, sineva2021negation}. To highlight this problem, \citet{garcia2023not} created a dataset to assess LLMs handling of negation in multiple NLP tasks. They found that LLMs struggle with negative sentences and lack a deep understanding of negation. The importance of negation in NLP tasks is also highlighted by \citet{hossain2022analysis}, which explores the role of negation in eight corpora for six popular natural language understanding tasks. The study finds that negation is virtually ignored by these corpora and that SotA transformers trained with these corpora face challenges in understanding negation. Unlike our work, all the mentioned works study negation solely in NLP tasks, and do not study their generalisation ability on ranking and IR-related tasks.

\header{Negation in Medical Information Retrieval.}  
Negation is a well-recognised challenge in medical IR, as clinical notes frequently state the \emph{absence} of conditions, which can mislead retrieval systems if not properly handled. Early rule-based tools like NegEx~\cite{chapman2001simple} and its extension ConText~\cite{harkema2009context} became standard components in clinical NLP pipelines, often integrated into IR systems such as those used in the TREC Medical Records track. These methods detect negation scopes around medical concepts and filter out negated mentions of query terms, significantly improving precision by reducing false positives. Subsequent machine-learning approaches—such as assertion classification in the i2b2 challenge~\cite{uzuner20112010} and cohort retrieval by \citet{taylor2018role} further advanced negation handling. These methods, along with early neural models like NegBio~\cite{peng2018negbio}, perform reasonably well in narrow clinical settings but tend to plateau beyond simple cue patterns and require extensive domain-specific tuning. However, in this work, we study various complex negation cases in both NevIR and ExcluIR, with a particular focus on how recent LLM-based architectures handle them, as well as their generalisability on passage ranking.

\header{Negation in Information Retrieval.}  
Negation handling has also been explored in broader IR contexts, often through rule-based or hybrid methods. For instance, \citet{goldin2003learning} demonstrated how negated terms can be filtered or reweighted in retrieval pipelines, showing that statistical models can effectively complement rule-based components. At the time of the NevIR paper's publication, little research had tackled negation in neural IR models directly. More recently, however, \citet{zhang2024excluir} introduced \emph{ExcluIR}, which provides an evaluation benchmark and training set aimed at improving how retrieval models handle exclusionary queries--negation expressed through user intent. While both NevIR and ExcluIR examine how IR models handle negation, they address different aspects of it: NevIR focuses on negation within documents, whereas ExcluIR emphasises the exclusionary nature of queries. \citet{malaviya2023quest} developed \emph{QUEST}, a dataset of natural language queries involving implicit set operations (including negation) that map to entities in Wikipedia. Their analysis aligns with findings from NevIR, showing that many modern retrieval models still struggle with negation-related phenomena. Unlike previous work, in our work we study the effect negation generalisation across two negation datasets (i.e., NevIR and ExcluIR), enabling us to study the generalisation capabilities of different ranking models.

\header{LLMs for Information Retrieval.}
As LLMs have demonstrated strong representational capabilities, they are increasingly used as backbones in retrieval and ranking architectures~\cite{zhularge}. These backbones are integrated into both bi-encoder and cross-encoder models, enhancing query-document representations for retrieval and reranking tasks~\cite{zhularge}. Around the introduction of NevIR~\cite{weller2024nevirnegationneuralinformation}, most LLM-based IR models primarily relied on what are now considered relatively smaller-scale models, such as \texttt{BERT}~\cite{BERT} and \texttt{T5}~\cite{T5}. With the emergence of larger and more accessible open-source LLMs, including those from \texttt{Llama-3}~\cite{llama3}, \texttt{Mistral}~\cite{mistral7b}, and \texttt{Qwen-2} \cite{yang2024qwen2technicalreport}, recent developments have integrated these models into IR systems, achieving state-of-the-art performance ~\cite{promptriever, repllama_rankllama, qwen_gte, gritlm}. With these improvements in LLM performance, researchers have also begun using LLMs directly as re-rankers, prompting them to reorder a list of retrieved documents given a query \cite{rankGPT, rankLLM}. \texttt{RankGPT}~\cite{rankGPT} demonstrated that generative LLMs, when prompted effectively, can outperform traditional re-rankers by leveraging their strong contextual understanding. Similarly, \texttt{RankLLM}~\cite{rankLLM} fine-tunes open-source LLMs for listwise document ranking, showing competitive performance with minimal additional training. To address the efficiency limitations of these models, \citet{FIRST}  introduce an approach that leverages the output logits of the first generated identifier to directly derive a rank order of the candidates. While NevIR evaluates the capabilities of LLM-based IR models for handling negation, we reassess the current SotA as the field is progressing rapidly. This includes addressing a key gap, as we are the first to evaluate the emerging class of listwise LLM re-rankers on their ability to handle negation.

% \moh{None of the existing work examines the capability of LLM-based ranking models in effectively ranking negation-based documents. In this work, we fill this research gap by examining the effectiveness of listwise LLM rankers on NevIR and ExcluIR.}

%% file: sections/03_nevir_overview.tex
\section{Overview of NevIR}
\begin{figure}[tbp]
  \centering
  \includegraphics[width=0.5\textwidth]{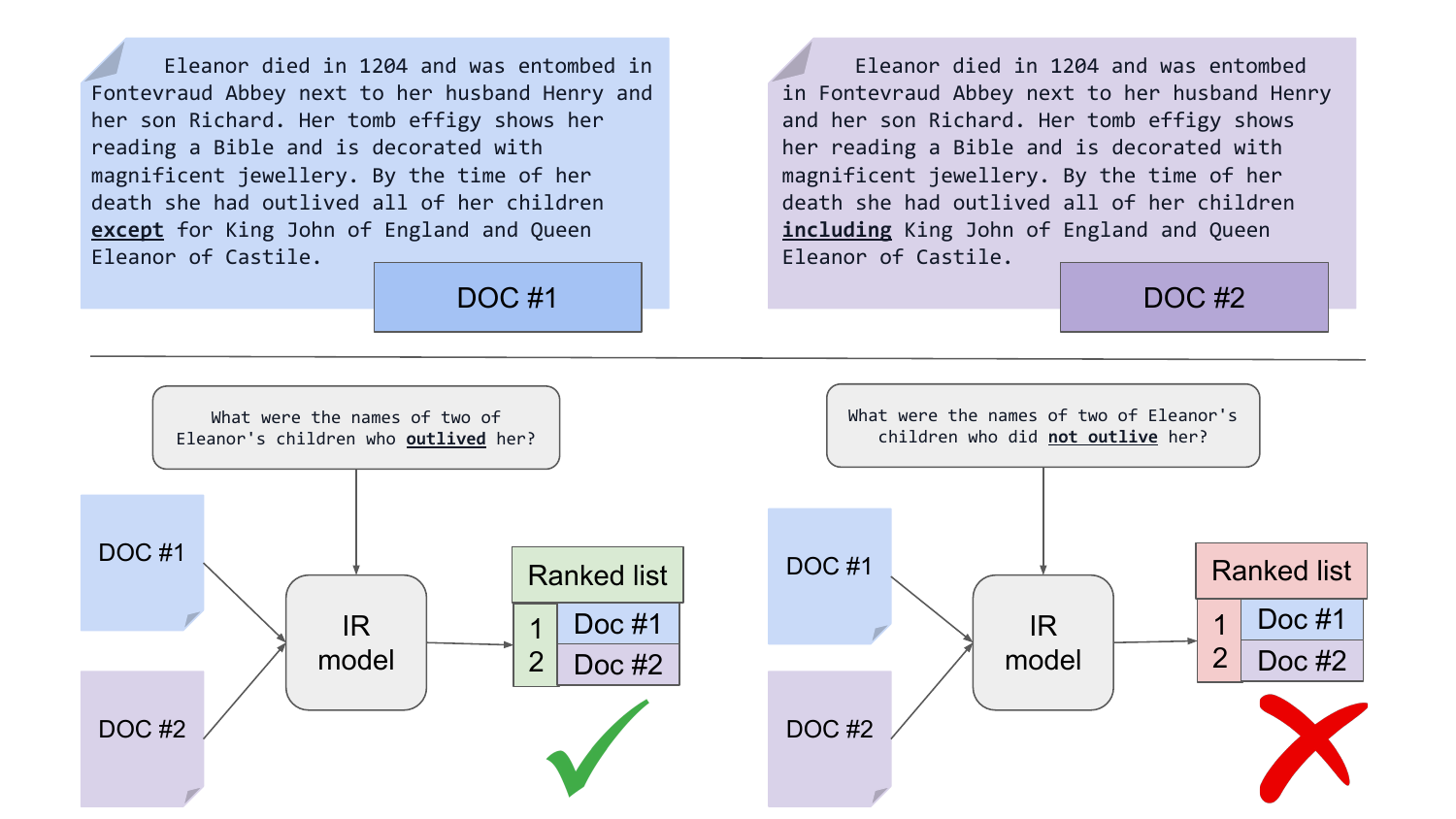}
  \caption{An example NevIR instance and the pairwise evaluation process for a contrastive query-document pair. This instance is classified as incorrect because only one of the two queries is ranked correctly, indicating that the ranker fails to account for negation.}
    \label{nevir_example}
\end{figure}

The NevIR dataset consists of contrastive query-document pairs where the documents are identical except for a key negation. Each query aligns semantically with one of the two documents, with Query \#1 corresponding to Doc \#1 and Query \#2 to Doc \#2. One such query-document pair is shown in \autoref{nevir_example}.

NevIR was constructed by first identifying pairs of minimally different documents that include negation, using the C\textsc{onda}QA dataset \cite{condaqa} as a foundation. C\textsc{onda}QA is an English reading comprehension dataset designed to facilitate the development of models capable of processing negation effectively. It consists of 14,182 question-answer pairs with over 200 unique negation cues. The dataset was created through crowdworkers paraphrasing passages by rewriting negated statements in a way that preserves the sentence's meaning. Building on top of this, NevIR introduced a query generation step via Amazon Mechanical Turk\footnote{\url{https//mturk.com}}. Crowdworkers were tasked with formulating one query per paragraph under specific constraints, ensuring that each query aligned most with one of the two minimally different documents.

For evaluation, a contrastive approach was employed that focuses on query-document pairs, rather than a typical IR corpus. Pairwise accuracy was used as the evaluation metric. This metric tests whether both ranked lists for the queries within the pair are accurate. The lower half of \autoref{nevir_example} shows the pairwise accuracy evaluation process. In this example, the IR model scored zero paired accuracy, ranking Doc \# 1 above Doc \# 2 in both queries (and failing to take into account the negation).

\header{Limitations.} 
A shortcoming of the original NevIR paper was that it relies on pairwise accuracy, a non-standard IR metric, limiting its broader applicability to IR tasks, such as ranking. In particular, the metric tests whether a model can differentiate negation between two given documents, offering insights into LMs' handling of negation. However, the experiments in the original paper suggest that pairwise accuracy does not correlate with ranking-based metrics, such as nDCG@5. Therefore, the original study mainly addressed negation from an NLP perspective, focusing solely on language understanding, while missing the IR aspect of it.

Another limitation of the paper was the absence of a baseline to assess the model's overall performance on the data distribution. While the authors included a random baseline (25\%) for the negation task, we argue that an additional baseline evaluating general performance on the task can provide more insights into the effectiveness of the experimented models. This would have helped distinguish whether the model's poor performance is due to challenges in ranking documents with negation or an Out-Of-Distribution effect.

%% file: sections/04_methodology.tex
\section{Experimental Setup}
Through the original code\footnote{\url{https://github.com/orionw/NevIR}} and valuable correspondence with the original authors--primarily focused on complementing the fine-tuning code--we reconstructed the experiments from the original work. In this section, we describe the details of the experimental setup for the reproducibility experiments and our extensions.

\subsection{Models} \label{model_description}
We evaluate nearly all the neural IR models\footnote{Due to dependency issues with the RocketQA library, it was not possible to reproduce the RocketQA models \cite{qu-etal-2021-rocketqa}} categorised and tested by \citet{weller2024nevirnegationneuralinformation}, as well as update the benchmark with newly developed SotA models. Below, we describe each category of retrieval models and discuss the models evaluated within each. Additionally, we introduce \textit{listwise LLM re-ranking} as a new model category, which emerged in recent work \cite{rankGPT, rankLLM, FIRST}.

\header{Random Baseline.}  
A baseline that randomly ranks each document in a pair. Since each pair has a 50\% chance of being ranked correctly, and there are two independent pairs, the expected mean pairwise accuracy is \(25\%\), calculated as \( \frac{1}{2} \times \frac{1}{2} = \frac{1}{4} \).

\header{Sparse Models and Learned Sparse Retrieval.}
Sparse IR models use bag-of-words representations for document retrieval. The original work evaluated traditional TF-IDF and two \texttt{SPLADEv2++} variants: the ensemble distillation and self-distillation methods~\cite{formal2022distillation, lassance2022efficiency}. To extend this category, we evaluated \texttt{SPLADE-v3} \cite{spladev3}, which introduces multiple negatives per batch during training, improved distillation scores, and additional fine-tuning steps.

\header{Late Interaction Models.}
Late interaction models encode queries and documents as vectors for each sub-word token. During inference, they compute similarity through a MaxSim operation. We reproduced \texttt{ColBERTv1}~\cite{colbertv1} and \texttt{ColBERTv2}~\cite{santhanam2021colbertv2}, and did not test any new models as these remain representative of the state of the art in this category.

\header{Bi-Encoders.}
Bi-encoders encode queries and documents into single-vector representations, enabling fast similarity computations using dot products or cosine similarity. The original work evaluated a variety of models, including some from SentenceTransformers \cite{sentencetransformers}, \texttt{DPR} \cite{dpr} and \texttt{CoCondenser}~\cite{gao-callan-2022-unsupervised}. We extended this category by evaluating several additional models. \texttt{DRAGON} \cite{dragon} is a model similar to \texttt{DPR} but adapted with advanced data augmentation techniques for enhanced generalisability. \texttt{Promptriever} \cite{promptriever} introduces promptable retrieval and was released with both \texttt{Llama-2-7B} and \texttt{Mistral-7B} backbones. We also included OpenAI's \texttt{text-embedding-small} and \texttt{text-embedding-large} models, and \texttt{RepLlama} \cite{repllama_rankllama}, which uses a \texttt{Llama-2} backbone. Furthermore, we evaluated the Qwen models with 1.5B and 7B variants \cite{yang2024qwen2technicalreport}, and \texttt{GritLM} \cite{gritlm}, which is based on a \texttt{Mistral-7B} backbone. \texttt{GritLM} is trained as a bi-encoder, cross-encoder, and generative model, but due to time constraints, we focused solely on its bi-encoder variant. Testing its cross-encoder and generative modes would be an interesting direction for future work.

\header{Cross-Encoders.}
Cross-encoders process both the document and the query together, offering highly expressive representations but at a higher computational cost. These models are typically used for re-ranking rather than retrieval. The original study evaluated various cross-encoders from SentenceTransformers~\cite{sentencetransformers} and MonoT5 \cite{nogueira-etal-2020-document}. In addition to these, we evaluated \texttt{RankLlama} \cite{repllama_rankllama}, trained as a pointwise re-ranker with a \texttt{Llama-2-7B} backbone. We also tested various BGE models, including \texttt{bge-reranker-base}\footnote{\url{https://huggingface.co/BAAI/bge-reranker-base}}, \texttt{bge-reranker-large}\footnote{\url{https://huggingface.co/BAAI/bge-reranker-large}}, and \texttt{bge-reranker-v2-m3}\footnote{\url{https://huggingface.co/BAAI/bge-reranker-v2-m3}}, which are parameter-efficient cross-encoders optimised for multilingual tasks. Finally, we included a Jina re-ranker, specifically \texttt{jina-reranker-v2-base-multilingual}\footnote{\url{https://huggingface.co/jinaai/jina-reranker-v2-base-multilingual}}, a re-ranking model designed for diverse cross-lingual retrieval tasks.

\header{Listwise LLM Re-rankers.}
In the listwise LLM re-ranking approach, LLMs, which are often pre-trained as foundation models, are adapted for ranking tasks. While these models are not explicitly fine-tuned for re-ranking in many cases, their rich contextual understanding and ability to assess semantic relevance enables them to reorder a given list of documents based on their alignment with a query. 
In this category, we evaluated \texttt{RankGPT}~\cite{rankGPT} and specifically tested \texttt{GPT-4o-mini}~\cite{openai2024gpt4technicalreport}, \texttt{GPT-4o}~\cite{openai2024gpt4technicalreport}, and \texttt{o3-mini}. Additionally, we included \texttt{Qwen2} models with both 1.5B and 7B Instruct variants \cite{qwen_gte}, as well as \texttt{Llama-3.2-3B-Instruct}, \texttt{Llama-3.1-7B-Instruct} \cite{llama3}, and \texttt{Mistral-7B-Instruct-v0.3} \cite{mistral7b}. These models span diverse parameter scales and are optimised for instruction-following tasks. 
For all LLMs, we employed standardised prompts from the \texttt{RankGPT} framework to ensure consistency across evaluations.

\subsection{Datasets and evaluation} \label{dataset_description} 
We evaluate the different families of models on several IR and negation datasets, namely:

\header{NevIR.} 
The NevIR \cite{weller2024nevirnegationneuralinformation} dataset is a benchmark for evaluating the impact of negation on neural information retrieval systems, featuring 2,556 contrastive document pairs and associated queries designed to test model understanding of negation. The dataset is divided into a train set (948 pairs, 37\%), a validation set (225 pairs, 9\%), and a test set (1,380 pairs, 54\%), with the test split being the largest. 

For the evaluation of NevIR we follow the setup of the original authors and use pairwise accuracy, illustrated in \autoref{fig:intro_example}. This metric assesses whether the model correctly flips the order of the ranking when given the negated query. 

\header{ExcluIR.} 
Unlike the NevIR dataset, which primarily focuses on understanding negation semantics within documents, ExcluIR emphasises the exclusionary nature of queries. The ExcluIR dataset \cite{zhang2024excluir} serves as both an evaluation benchmark and a training set designed to help retrieval models comprehend exclusionary queries. The training set contains 70,293 exclusionary queries generated by \texttt{ChatGPT-3.5}, each paired with a positive and a negative document. The evaluation benchmark consists of 3,452 manually annotated exclusionary queries. For our fine-tuning experiments, we use the benchmark dataset, which is divided into a train set (2,070 pairs, 60\%), a validation set (346 pairs, 10\%), and a test set (1,040 pairs, 30\%). 

Each sample in the ExcluIR evaluation dataset consists of a single exclusionary query and two documents. The dataset is designed so that the model should rank one document higher than the other. We use accuracy as the evaluation metric.

\header{MS MARCO.} 
The MS MARCO dataset \cite{bajaj2016ms} is a large-scale benchmark for information retrieval and question answering, containing millions of real-world queries and corresponding passages. Designed to evaluate retriever and reader systems in a general IR setting, it includes only minimal instances of negation. Due to the large size of the MS MARCO development set, we chose to only use a subset of the original development set for evaluation, following a similar setting to TAS-B \cite{hofstatter2021efficiently}. This subset contained 500 queries, following the approach of SPLADE \cite{formal2021splade}. For every query, we used the top 1000 relevant documents as ranked by bm25 to build a corpus. 

For evaluating retrieval effectiveness on a large-scale IR dataset like MS MARCO, we use MRR@10.

\subsection{Listwise LLM Re-rankers}
Unlike other model categories that rely on triplets for fine-tuning, LLMs require conversational datasets. To facilitate this, we converted NevIR and ExcluIR to datasets with a conversational structure and uploaded them to Hugging Face\footnote{\url{https://huggingface.co/datasets/thijmennijdam/NevIR-conversational},\\ \- \url{https://huggingface.co/datasets/thijmennijdam/ExcluIR-conversational}}. The dataset construction follows the instructional permutation generation method from \texttt{RankGPT}~\cite{rankGPT}, enabling models to directly output a ranked list given a set of candidate passages. This approach involves presenting a group of passages, each assigned a unique identifier, to the LLM, which is then prompted to generate their permutation in descending order of relevance to the query.

\header{Sliding Window Strategy.} 
To evaluate the Listwise LLM re-rankers on MS MARCO, we addressed the token limitations of LLMs by adopting a sliding window strategy \cite{rankGPT}. This strategy introduces two hyperparameters: window size and step size. Due to computational constraints, the number of documents that can be re-ranked is limited. Therefore, we restricted re-ranking to the top 30 documents retrieved by BM25~\cite{bm25}. A problem we encountered was that fine-tuning to re-rank more than two documents at a time (e.g., setting the window size higher than two) on the MS MARCO dataset led to a decline in ranking performance when re-ranking only the two documents in NevIR. To prevent this issue from impacting evaluation on NevIR, we set the sliding window size to 2 and the step size to 1 in all our fine-tuning experiments.

\header{Few-shot Prompting.} 
To assess the generalisability of LLMs to negation, we evaluated their ability to learn from examples through few-shot learning with 1, 3, and 5 examples. For each query in the NevIR test set, we provided the LLM with a fixed set of examples from the NevIR training set. This ensured that the LLMs were exposed to the same training samples for every test query.

\subsection{Experimental Details}

\header{Hyperparameters.}
For fine-tuning the different ranking models, we used their respective default hyperparameters for training, as recommended in their original papers \cite{colbertv1, nogueira-etal-2020-document, reimers-gurevych-2019-sentence}. Specifically, the ColBERT models use a default learning rate of 3e-6, the bi-encoder model from Sentence-Transformers uses 2e-5, and the cross-encoder model \texttt{MonoT5-base} employs a learning rate of 3e-4. For the \texttt{Mistral-7B-Instruct-v0.3} model, we use a learning rate of 2e-4.

\header{Computational Requirements.}
Our experiments were conducted using an NVIDIA H100 GPU. It takes approximately 50 hours to reproduce our experiments. Additionally, we spent \$40 on OpenAI API costs.

%% file: sections/05_experimental_results.tex
\section{Experimental Results}
In this section, we present and describe the results of our reproducibility experiments as well as those of the extensions.

\subsection{Negation Understanding on SotA Neural Architectures}
The results of reproducing the NevIR benchmark are presented in \autoref{tab:model_comparison}. Overall, the reproduced scores align closely with those reported in the original paper, with only minor deviations observed for the \texttt{SPLADE}, \texttt{ColBERT}, and \texttt{DPR} models.

All models originally evaluated exhibit poor performance, with scores significantly below the human baseline of 100\% \cite{weller2024nevirnegationneuralinformation}. This prompts the first research question: 
\begin{enumerate}[label=\textbf{RQ\arabic*}]
    \item How do the negation understanding results of the NevIR dataset generalise to the most recent retrieval architectures, such as listwise LLM re-rankers?\label{rq1}
\end{enumerate}

\input{tables/table1}

\header{Evaluation of the SotA on NevIR.}
\autoref{tab:model_comparison} presents the results of benchmarking newly developed SotA models. Across all categories, we observe that at least one new SotA model demonstrates improvements on the NevIR benchmark over previously tested models.

Regarding the bi-encoders, the \texttt{GritLM-7B} model achieves significant improvements (39\%), outperforming other models in this category by a substantial margin. In contrast, models such as \texttt{RepLlama}, \texttt{PromptRetriever} variants, and \texttt{gte-Qwen-2} variants show comparatively minor improvements over previously tested bi-encoders, despite utilising larger LLM backbones such as \texttt{Llama-2}, \texttt{Qwen} (1.5B and 7B), and \texttt{Mistral-7B}. Notably, \texttt{GritLM} also employs a \texttt{Mistral-7B} backbone and functions as both an embedding and generative model, potentially enhancing its negation understanding compared to traditional embedding models. This strong performance of \texttt{GritLM} suggests that bi-encoders may eventually develop more effective representations for negation understanding. However, further analysis is needed to identify the specific factors contributing to this superior performance.

In the cross-encoder category, the integration of \texttt{Llama-7B} in \texttt{RankLlama} (31.6\%) does not improve performance over previous SotA models. The various BGE re-rankers we benchmarked also show only limited improvement compared to earlier evaluations. In contrast, the \texttt{jina-reranker-v2-base-multilingual} model achieves a pairwise accuracy of 65.2\% and significantly outperforms all other cross-encoders tested, despite its smaller model size. Since both the various BGE re-rankers and the \texttt{jina-reranker-v2-base-multilingual} are trained on multilingual data, we hypothesise that this performance difference is due to the \texttt{jina-reranker-v2-base-multilingual} having been trained on a larger amount or higher quality negation data.

\header{Listwise LLM Re-rankers.}
In the newly introduced category of listwise LLM re-rankers, a clearer relationship emerges between model scale and performance, with larger models generally achieving better results. However, performance varies significantly among models with 7B parameters. For instance, while the \texttt{Llama 3.1-7B} model performs near random, the \texttt{Qwen2-7B-Instruct} (36.9\%) and \texttt{Mistral-7B-Instruct} (46.3\%) models demonstrate substantially stronger performance. Scaling further, \texttt{GPT-4o-mini} (64.1\%) and \texttt{GPT-4o} (70.1\%) illustrate the benefits of increased model capacity, as parameter estimates suggest these models are considerably larger. Notably, \texttt{o3-mini}, which incorporates reasoning capabilities, improves performance even further to 77\%.

\input{tables/table2}

We analysed errors made by the top-performing models; \texttt{GPT-4o} and \texttt{o3-mini}. This qualitative assessment suggests that their mistakes often occur on more challenging samples. For example, both made a mistake when the query asks whether the event was officially organised, and the documents distinguish between invitations to \textit{formal} and \textit{informal exhibitions}. This task not only necessitates an understanding of negation but also of real-world knowledge about formal events, highlighting the inherent difficulty of achieving a perfect score on the NevIR dataset.

We answer \ref{rq1} by finding that, while models have improved across all categories, negation understanding remains a challenge. Although the listwise LLM re-rankers achieve the highest performance on the benchmark, their practical use is limited by significantly higher computational demands.

% \vspace{-0.2em}
\subsection{Results on the Negation Learning Experiments}

\begin{figure}[tbp]
  \centering
  \includegraphics[width=0.5\textwidth]{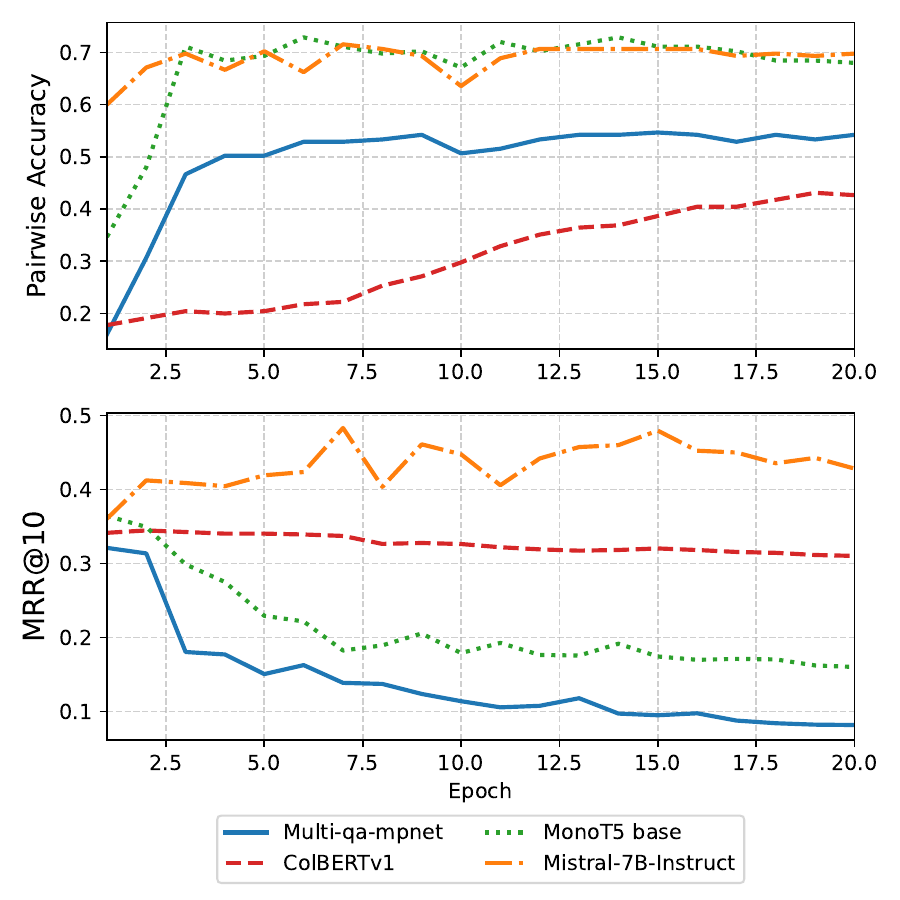}
  \caption{Fine-tuning results on NevIR. The top plot shows pairwise accuracy on NevIR, while the bottom plot presents MRR@10 on MS MARCO.}
  \label{fig:nev_fine_tune}
\end{figure}

\header{Reproducing fine-tuning on NevIR.}
The results of the reproduced fine-tuning experiment on NevIR are presented in \autoref{fig:nev_fine_tune}. Consistent with the original paper, all models show improvements in pairwise accuracy after fine-tuning, with the bi-encoder and late interaction models reaching around 50\%, and the cross-encoder achieving approximately 75\%. However, none of the models attain perfect accuracy, indicating potential for further improvement. This led us to consider how listwise LLM re-rankers perform when fine-tuned on NevIR data, prompting our second research question:

\begin{enumerate}[label=\textbf{RQ\arabic*}, start=2]
    \item How do listwise LLM re-rankers perform when fine-tuned on the negation data of NevIR?\label{rq2}
\end{enumerate}

\header{Fine-tuning listwise LLM re-ranker on NevIR.}
In \autoref{fig:nev_fine_tune} we also include the results of fine-tuning a listwise LLM re-ranker, \texttt{Mistral-7B-Instruct-v0.3}, on NevIR. We observe a significant improvement in negation handling over the base performance of 46.3\% reported in \autoref{tab:model_comparison}. However, the NevIR score after fine-tuning is comparable to that of the cross-encoder, indicating that, despite its superior base performance, this model exhibits similar learning capabilities for the negation task as the more traditional cross-encoder model.

\header{Few-shot listwise LLM-reranker on NevIR.}
\autoref{tab:few_shot} shows the results of the few-shot experiments. Notably, the performance of \texttt{Mistral-7B-Instruct-v0.3} drops when going from zero-shot to few-shot learning. This suggests that the model struggles to internalise and generalise the provided few-shot negation patterns. In contrast, \texttt{GPT-4o} demonstrates improved performance with the inclusion of more examples, indicating a positive trend in few-shot learning capabilities. These findings align with previous research indicating that larger LMs tend to perform better in few-shot learning scenarios~\cite{brown2020languagemodelsfewshotlearners}.

This partly answers \ref{rq2}, fine-tuning on NevIR can significantly enhance performance for a listwise LLM re-ranker, suggesting an improved ability to handle negation. However, this conclusion remains uncertain as there may be other contributing factors, such as the model learning a shortcut, e.g. the data distribution, to solve the task. This led us to consider the third research question: 
\begin{enumerate}[label=\textbf{RQ\arabic*}, start=3]
    \item How does fine-tuning on other negation datasets, such as ExcluIR, improve generalisability on NevIR?\label{rq3}
\end{enumerate}

\subsection{Examining Patterns of Generalised Negation Comprehension}
\begin{figure}[tbp]
  \centering
  \includegraphics[width=0.48\textwidth]{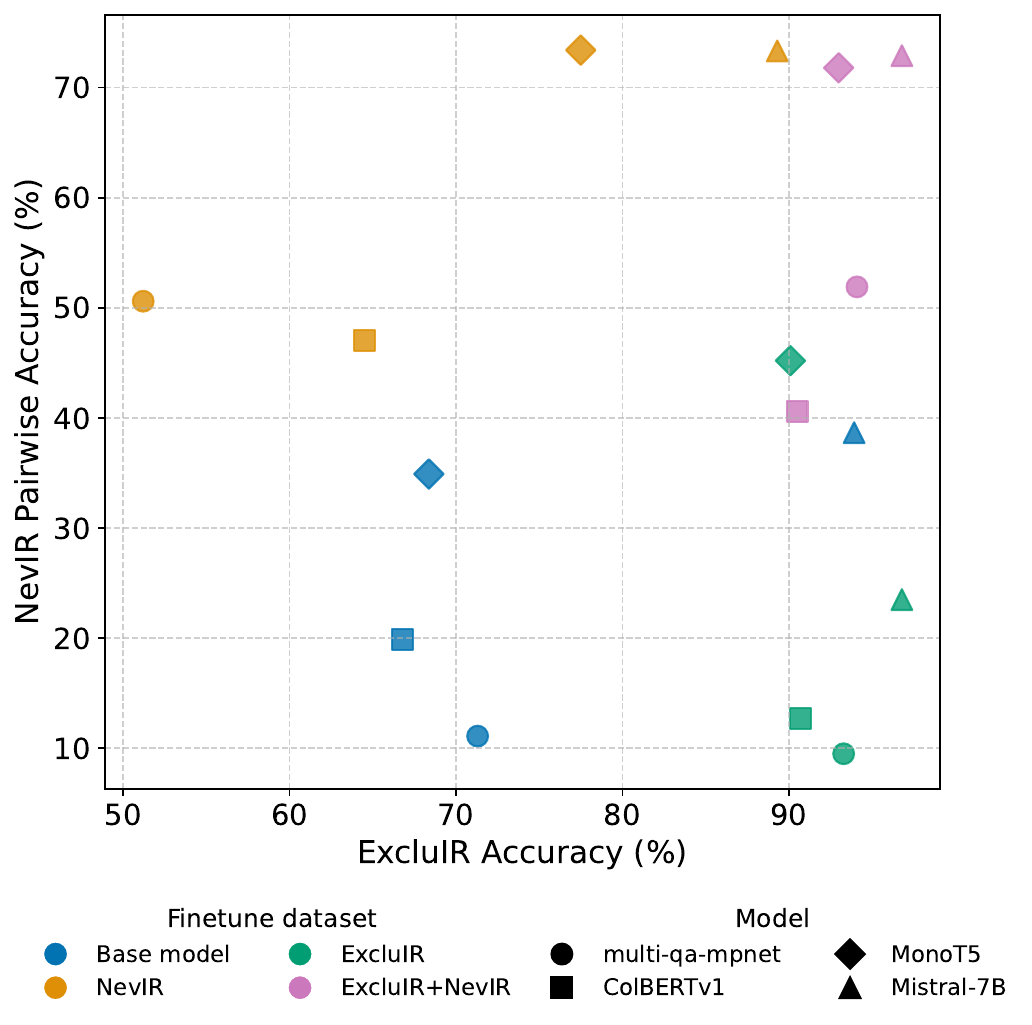}
  \caption{Model performance on ExcluIR and NevIR of the four families of models. Models were fine-tuned on each dataset separately and on the merged dataset.}
   \label{fig:merge_test_performance}
\end{figure}

To test how well negation comprehension learned from the NevIR dataset generalises to different negation scenarios, we fine-tune various IR models on NevIR and evaluate on a different negation dataset, ExcluIR. To assess generalisation fully, we also fine-tune models on ExcluIR and evaluate on NevIR. The results are presented in \autoref{fig:merge_test_performance}.

For the bi-encoder, late interaction and listwise LLM re-ranker models, the ExcluIR performance does not improve when fine-tuned on the NevIR dataset and only shows an increase when the ExcluIR data is included in the training set. Similarly, the NevIR performance also does not improve when fine-tuned on ExcluIR. This shows that, for these models, the seeming negation comprehension does not transfer effectively to other negation scenarios. 

In contrast, for the \texttt{MonoT5-base} model, we observe that fine-tuning on the NevIR dataset leads to an improvement in the ExcluIR score as well. A similar trend is observed when fine-tuning on ExcluIR and evaluating on NevIR. This indicates that  \texttt{MonoT5-base} is capable of generalising its negation comprehension to a different dataset, due to its large model size and the expressive representation of cross-encoders.

Additionally, we test all models on a merged dataset where the training sets of NevIR and ExcluIR are combined. Across all models, this results in either the highest or close to the highest accuracy on both datasets, indicating that training on both datasets simultaneously leads to the best overall performance. 

We answer \ref{rq3} by finding that NevIR and ExcluIR may capture distinct aspects of negation that require explicit exposure during training to be effectively learned. As a result, models show limited to no generalisability across these datasets.

\subsection{Analysis of Training Behaviour}
The original study found that certain models exhibited overfitting to negation data during fine-tuning. While this improved their performance on negation-related queries, it came at the expense of their general ranking capabilities. This trade-off motivated the formulation of the fourth research question:

\begin{enumerate}[label=\textbf{RQ\arabic*}, start=4]  
    \item How can performance on a negation dataset be improved while maintaining strong ranking capabilities? \label{rq4}  
\end{enumerate}

\autoref{fig:nev_fine_tune} presents the learning curves for NevIR and MS MARCO during fine-tuning on NevIR. Consistent with findings in the original paper, we observe a decline in MRR, particularly in bi-encoder and cross-encoder models, namely \texttt{multi-qa-mpnet-base-dot-v1} and \texttt{MonoT5-base}. Their steeper learning curves show that these models may be learning faster, but are also more prone to overfitting. Additionally, we evaluate the \texttt{Mistral-7B-Instruct-v0.3} listwise re-ranker model on MS MARCO using a sliding window approach. Unlike the other models, \texttt{Mistral-7B-Instruct-v0.3} maintains a stable MRR throughout fine-tuning, indicating that learning to re-rank on NevIR does not inherently degrade its ability to rank documents on MS MARCO.

\begin{figure}[tbp]
    \centering
    \includegraphics[width=0.5\textwidth]{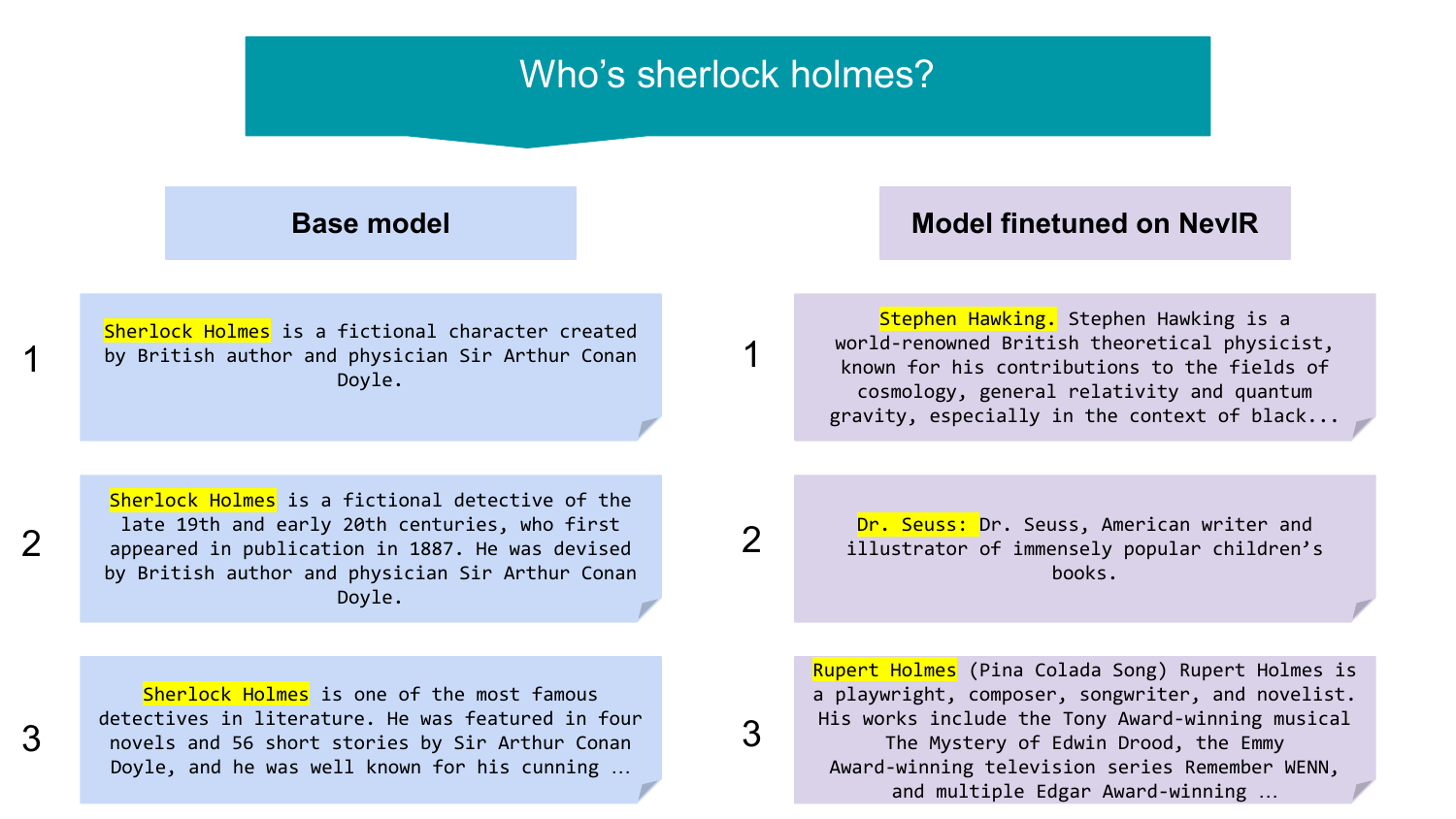}
    \caption{An MS MARCO query with the rankings by the \texttt{multi-qa-mpnet} model and its fine-tuned variant on NevIR.}
    \label{fig:msmarco_analysis}
\end{figure}

To further investigate the impact of overfitting,  we compare the ranking produced by the original \texttt{multi-qa-mpnet-base-dot-v1} model and the variant fine-tuned on NevIR. We observed that after fine-tuning, the model fails at exact term matching, instead favouring documents featuring other well-known individuals. An example highlighting this is provided in \autoref{fig:msmarco_analysis}. This suggests that learning a new task overrides fundamental ranking capabilities, possibly because the model shifts toward contrastive learning, emphasising negation handling over traditional lexical matching.

\input{tables/table4}

We find that the effects of overfitting can be partially mitigated by adopting a more effective early stopping criterion. Rather than selecting the best model based solely on NevIR validation performance, we apply a trade-off criterion that optimises for both NevIR and MS MARCO performance and select the optimal model based on the average of MRR@10 and pairwise accuracy. \autoref{tab:selection_criterium} shows that this trade-off selection preserves NevIR performance while significantly reducing degradation on MS MARCO, thus mitigating overfitting.

We answer \ref{rq4} by finding that overfitting to negation data can degrade passage ranking capabilities, but a trade-off-based hyperparameter selection criterion helps maintain performance on both negation and MS MARCO.

%% file: tables/table1.tex
\begin{table*}[tbp]
  \caption{Model performance comparison on NevIR in terms of pairwise accuracy (P.\ Acc). We include both newly tested and reproduced models, showing the original and our reproduced scores where available. An ``X'' indicates cases where no original score was reported. Training dataset indicates which dataset the models are trained on from scratch.}
  \label{tab:model_comparison}
  \begin{tabular}{llllcc}
    \toprule
    Type & Training Dataset & Params & Model Name & P.\ Acc & P.\ Acc (ours) \\
    \midrule
    Random & N/A & 0 & Random & 25\% & 25\%\\
    \midrule
    Sparse & N/A & N/A & TF-IDF & 2.0\% & 2.0\% \\
    & MS MARCO & 110M & SPLADEv2 ensemble-distill \cite{formal2022distillation} & 8.0\% & 7.9\%\\
    & MS MARCO & 110M & SPLADEv2 self-distill \cite{formal2022distillation} & 8.7\% & 8.5\% \\
    & MS MARCO & 110M & SpladeV3 self-distill \cite{spladev3} & X & \textbf{9.9\%} \\
    \midrule
    Late Interaction & MS MARCO & 110M & ColBERTv2 \cite{santhanam2021colbertv2} & 13.0\% & 12.8\% \\
    & MS MARCO & 110M & ColBERTv1 \cite{colbertv1}& 19.7\% & \textbf{19.8\%}\\
    \midrule
    Bi-Encoders & NQ & 219M & DPR \cite{dpr} & 6.8\% & 6.5\%\\
    & MS MARCO & 110M & MS MARCO-bert-base-dot-v5 & 6.9\% & 6.9\%\\
    & MS MARCO & 110M & coCondenser \cite{gao-callan-2022-unsupervised} & 7.7\% & 7.7\%\\
    & NQ & 66M & nq-distilbert-base-v1 & 8.0\% & 8.0\%\\
    & MS MARCO & 110M & all-mpnet-base-v2 & 8.1\% & 8.1\%\\
    & MS MARCO & 66M & MS MARCO-distilbert-cos-v5 & 8.7\% & 8.7\%\\
    & QA Data & 110M & multi-qa-mpnet-base-dot-v1 & 11.1\% & 11.1\%\\
    & MS MARCO & 110M & DRAGON \cite{dragon} & X & 6.8\% \\
    & Variety of data sources \cite{qwen_gte}  & 1.5B & gte-Qwen2-1.5B-instruct \cite{qwen_gte} & X & 11.8\% \\
    & MS MARCO & 7B & RepLlama \cite{repllama_rankllama} & X & 13.0\%  \\
    & Unavailable & Unavailable & OpenAI text-embedding-3-small & X & 16.2\% \\
    & MS MARCO & 8B\footnotemark & promptriever-llama3.1-8b-v1 \cite{promptriever} & X & 18.1\% \\
    & Variety of data sources \cite{qwen_gte}  & 7B & gte-Qwen2-7B-instruct \cite{qwen_gte} & X & 19.0\% \\
    & MS MARCO & 7B & promptriever-mistral-v0.1-7b-v1 \cite{promptriever} & X & 19.7\% \\
    & Unavailable & Unavailable & OpenAI text-embedding-3-large & X & 22.6\%\\
    & E5 \cite{e5} \& Tülu 2 \cite{tulu2} & 7B & GritLM-7B \cite{gritlm} & X & \textbf{39.0\%} \\
    \midrule
    Cross-Encoders & STSB & 355M & stsb-roberta-large & 24.9\% & 24.9\%\\
    & MS MARCO & 61M & MonoT5-small \cite{nogueira-etal-2020-document} & 27.7\% & 27.7\%\\
    & MNLI & 184M & nli-deberta-v3-base & 30.2\% & 30.2\%\\
    & QNLI & 110M & qnli-electra-base & 34.1\% & 34.1\%\\
    & MS MARCO & 223M & MonoT5-base \cite{nogueira-etal-2020-document} & 34.9\% & 34.9\%\\
    & MS MARCO & 2.85B & MonoT5-3B \cite{nogueira-etal-2020-document} & 50.6\% & 50.6\%\\
    & MS MARCO & 7B & RankLlama \cite{repllama_rankllama} & X & 31.6\% \\
    & Unavailable & 270M & bge-reranker-base & X & 32.2\% \\
    & Unavailable & 560M & bge-reranker-large & X & 42.7\% \\
    & Unavailable & 568M & bge-reranker-v2-m3 \cite{baai_bge_reranker_v2_m3} & X & 43.5\% \\
    & Unavailable & 278M & jina-reranker-v2-base-multilingual & X & \textbf{65.2\%} \\
    \midrule
    Listwise LLM Re-ranking & Variety of data sources \cite{qwen_gte} & 1.5B & Qwen2-1.5B-Instruct \cite{qwen_gte} & X & 11.4\% \\
    & Variety of data sources \cite{llama3} & 3B & Llama-3.2-3B-Instruct \cite{llama3} & X & 25.7\% \\
    & Variety of data sources \cite{llama3} & 7B & Llama-3.1-7B-Instruct \cite{llama3} & X & 29.7\% \\
    & Variety of data sources \cite{qwen_gte} & 7B & Qwen2-7B-Instruct \cite{qwen_gte} & X & 36.9\% \\
    & Variety of data sources \cite{mistral7b} & 7B & Mistral-7B-Instruct-v0.3 \cite{mistral7b} & X & 46.3\% \\
    & Unavailable & Unavailable & RankGPT 4o-mini \cite{openai2024gpt4technicalreport} & X & 64.1\% \\
     & Unavailable & Unavailable & RankGPT 4o \cite{openai2024gpt4technicalreport} & X & 70.1\% \\
     & Unavailable & Unavailable & RankGPT o3-mini & X & \textbf{77.3\%} \\
    \bottomrule
  \end{tabular}
\end{table*}

%% file: tables/table2.tex
\begin{table}[tbp]
  \caption{Performance of \texttt{GPT-4o} and \texttt{Mistral-7B-Instruct-v0.3} as listwise LLM re-rankers on the NevIR test set with few-shot prompting.}
  \label{tab:few_shot}
  \centering
  \begin{tabular}{llc}
    \toprule
    \textbf{Model Name} & \textbf{\# Shots} & \textbf{Pairwise Accuracy} \\
    \midrule
    \multirow{4}{*}{GPT-4o} & Zero-shot & 70.1\% \\
                             & 1-shot    & 72.0\% \\
                             & 3-shot    & 74.5\% \\
                             & 5-shot    & 76.9\% \\
    \midrule
    \multirow{4}{*}{Mistral-7B-Instruct-v0.3} & Zero-shot & 46.3\% \\
                                              & 1-shot    & 42.6\% \\
                                              & 3-shot    & 37.1\% \\
                                              & 5-shot    & 39.0\% \\
    \bottomrule
  \end{tabular}
\end{table}

    % GPT 4o-mini & zero-shot & 64.1\\
    % GPT 4o-mini & 1-shot & 59.2\\
    % GPT-4o-mini & 3-shot & 58.5 (first 3) ->  62.5 (last 3)\\
    % GPT 4o-mini & 5-shot & 63.1\\
    
% \begin{table}[h]
%   \centering
%   \caption{Model performance of listwise LLM models on NevIR when fine-tuned with varying types of few-shot learning.}
%   \label{tab:few_shot}
%   \begin{tabular}{lcc}
%     \toprule
%     Model & Finetune Set & NevIR Score \\
%     \midrule
%     GPT-4o  & Zero-shot & 70.1 \\
%     GPT-4o  & 1-shot & 72.0 \\
%     GPT-4o  & 3-shot & 74.5 \\
%     GPT-4o  & 5-shot & 76.9 \\
%     \midrule
%     Mistral-7B-Instruct-v0.3  & Zero-shot & 46.3 \\
%     Mistral-7B-Instruct-v0.3  & 1-shot & XXX \\
%     Mistral-7B-Instruct-v0.3  & 3-shot & 37.1 \\
%     Mistral-7B-Instruct-v0.3  & 5-shot & XXX \\
%     \bottomrule
%   \end{tabular}
% \end{table}

%% file: tables/table4.tex
\begin{table}[tbp]
\caption{Model performance on NevIR and MS MARCO under different selection criteria. The NevIR criterion selects the best model based on validation performance on NevIR, while the trade-off criterion balances NevIR performance with MS MARCO MRR@10 by taking best average performance.}
\resizebox{\columnwidth}{!}{%
\begin{tabular}{lccc}
\toprule
Model                      & Selection criterion & NevIR P. Acc & MSMarco MRR@10 \\ \midrule
multi-qa-mpnet-base-dot-v1 & NevIR               & 50.1\%     & 0.06       \\
multi-qa-mpnet-base-dot-v1 & Trade-off           & 48.9\%     & 0.20       \\
MonoT5-base                & NevIR               &  71.3\%           &   0.16             \\
MonoT5-base                & Trade-off           &  71.6\%           & 0.30               \\ \bottomrule
\end{tabular}%
}

\label{tab:selection_criterium}
\end{table}

%% file: sections/06_discussion_conclusions.tex
\vspace{-0.1em}
\section{Discussion and Conclusion}
We reproduce and extend the findings of NevIR, a benchmark study that
revealed most ranking models perform at or below the level of random ranking when dealing with negation. We consider several key factors: (i) the ongoing development of neural IR retrieval models; (ii) the emergence of a new family of listwise re-rankers based
on LLMs; and (iii) the emergence of new datasets on negation, like ExcluIR, enabling one to measure the generalisability of models in terms of negation understanding.

In our reproducibility experiments, we confirm that most tested models fail to account for negation. We also reproduce the finding that fine-tuning on negation-specific data leads to performance gains on NevIR but can also cause overfitting, resulting in reduced performance on general retrieval tasks such as MS MARCO. Through further qualitative analysis, we show that overfitting on NevIR overrides fundamental ranking heuristics, such as exact term matching. To mitigate this effect, we apply a selection criterion for early stopping.

Beyond reproducing the results of the original paper, we extend the experiments by evaluating newly developed SotA ranking models. We observe improvements across all tested categories, with notable advancements in bi-encoders and cross-encoders. Specifically, \texttt{GritLM}, a bi-encoder model, demonstrates substantial improvements, achieving performance comparable to many cross-encoders. Similarly, in the cross-encoder category, \texttt{jina-reranker-v2-base-multilingual} outperforms \texttt{MonoT5-3B}, despite having a significantly smaller model size.

Also, we tested the latest listwise LLM re-rankers on the NevIR task and found that they achieved the highest performance on NevIR, benefiting significantly from scaling. Furthermore, our results suggest that the reasoning capabilities of LLMs enhance performance, as \texttt{o3-mini} outperforms \texttt{GPT-4o} considerably. However, these models come at a significant computational cost, limiting their practicality to final-stage re-ranking. Additionally, we fine-tuned \texttt{Mistral-7B-Instruct-v0.3} and observed improved performance on NevIR. However, unlike \texttt{GPT-4o}, \texttt{Mistral-7B-Instruct-v0.3}'s performance degrades in few-shot learning, where \texttt{GPT-4o} has substantial gains. This indicates that only \texttt{GPT-4o} is capable enough to effectively grasp context and improve negation understanding.

To investigate the generalisation of negation comprehension learned from NevIR, we fine-tuned the models on NevIR and evaluated them on ExcluIR, and vice versa. The results show that only the cross-encoder model successfully transferred its negation understanding to the new setting. We attribute this to the cross-encoder's larger model size and its ability to combine the query and documents into a more expressive joint representation compared to other model categories. We find that the cross-encoder model captures negation in a way that meaningfully generalises across different data distributions and contexts.

Despite the positive trend we observe on the NevIR benchmark, our findings show that many neural IR models still struggle with negation. Even the best models continue to fall short of human-level performance, reinforcing that negation understanding remains a persistent challenge for neural IR systems.

\section*{Limitations}
One limitation of our assessment of re-ranking with the Mistral model is that we trained it on triplets from NevIR rather than on a list of documents, as NevIR does not contain listwise training data. As a result, we could not evaluate performance degradation on MS MARCO with a sliding window larger than two, since fine-tuning the model on lists longer than two made it unable to correctly handle the re-ranking task on NevIR.

Additionally, while the Mistral model seems to perform well on re-ranking MS MARCO, our evaluation was limited to 87 queries that contained at least one relevant document retrieved by BM25 within the top 30 results, as listwise LLM re-rankers depend heavily on the quality of the first-stage retrieval.

Finally, due to computational constraints, we were unable to evaluate and fine-tune an LLM larger than 7B parameters. A larger model could provide further insights into how model size affects re-ranking performance.

\subsection*{Environmental Impact}
The total GPU usage for this research, including exploratory and repeated runs, was approximately 100 hours on NVIDIA H100 PCIe GPUs (TDP: 350W), using the Snellius supercomputer (carbon efficiency: 0.421 kgCO$_2$eq/kWh).

We estimated emissions post hoc using the \href{https://mlco2.github.io/impact#compute}{MachineLearning Impact calculator} \cite{lacoste2019quantifying}, selecting an A100 PCIe 40/80GB GPU. Total emissions were approximately 13.16 kgCO$_2$eq, accounting for GPU power use and infrastructure efficiency.

Experiments using ChatGPT models were conducted via the OpenAI API. These models require substantial computing and water resources. Although exact figures are unavailable, we highlight their environmental cost and encourage future research to account for such impacts.

%% file: sections/07_acknowledgments.tex
\begin{acks}

This research was in part supported by the Swiss National Science Foundation (SNSF), under the project PACINO (Personality And Conversational INformatiOn Access), grant number 215742, and in part supported by the Informatics Institute (IvI) of the University of Amsterdam.

\end{acks}

%% file: main.bbl

\begin{thebibliography}{51}

%%% the Web address.

\ifx \showCODEN    \undefined \def \showCODEN     #1{\unskip}     \fi
\ifx \showISBNx    \undefined \def \showISBNx     #1{\unskip}     \fi
\ifx \showISBNxiii \undefined \def \showISBNxiii  #1{\unskip}     \fi
\ifx \showISSN     \undefined \def \showISSN      #1{\unskip}     \fi
\ifx \showLCCN     \undefined \def \showLCCN      #1{\unskip}     \fi
\ifx \shownote     \undefined \def \shownote      #1{#1}          \fi
\ifx \showarticletitle \undefined \def \showarticletitle #1{#1}   \fi
\ifx \showURL      \undefined \def \showURL       {\relax}        \fi
% invisible to TeX
\providecommand\bibfield[2]{#2}
\providecommand\bibinfo[2]{#2}
\providecommand\natexlab[1]{#1}
\providecommand\showeprint[2][]{arXiv:#2}

\bibitem[Bajaj et~al\mbox{.}(2016)]%
        {bajaj2016ms}
\bibfield{author}{\bibinfo{person}{Payal Bajaj}, \bibinfo{person}{Daniel Campos}, \bibinfo{person}{Nick Craswell}, \bibinfo{person}{Li Deng}, \bibinfo{person}{Jianfeng Gao}, \bibinfo{person}{Xiaodong Liu}, \bibinfo{person}{Rangan Majumder}, \bibinfo{person}{Andrew McNamara}, \bibinfo{person}{Bhaskar Mitra}, \bibinfo{person}{Tri Nguyen}, {et~al\mbox{.}}} \bibinfo{year}{2016}\natexlab{}.
\newblock \showarticletitle{Ms marco: A human generated machine reading comprehension dataset}.
\newblock \bibinfo{journal}{\emph{arXiv preprint arXiv:1611.09268}} (\bibinfo{year}{2016}).
\newblock


\bibitem[Brown et~al\mbox{.}(2020)]%
        {brown2020languagemodelsfewshotlearners}
\bibfield{author}{\bibinfo{person}{Tom~B. Brown}, \bibinfo{person}{Benjamin Mann}, \bibinfo{person}{Nick Ryder}, \bibinfo{person}{Melanie Subbiah}, \bibinfo{person}{Jared Kaplan}, \bibinfo{person}{Prafulla Dhariwal}, \bibinfo{person}{Arvind Neelakantan}, \bibinfo{person}{Pranav Shyam}, \bibinfo{person}{Girish Sastry}, \bibinfo{person}{Amanda Askell}, \bibinfo{person}{Sandhini Agarwal}, \bibinfo{person}{Ariel Herbert-Voss}, \bibinfo{person}{Gretchen Krueger}, \bibinfo{person}{Tom Henighan}, \bibinfo{person}{Rewon Child}, \bibinfo{person}{Aditya Ramesh}, \bibinfo{person}{Daniel~M. Ziegler}, \bibinfo{person}{Jeffrey Wu}, \bibinfo{person}{Clemens Winter}, \bibinfo{person}{Christopher Hesse}, \bibinfo{person}{Mark Chen}, \bibinfo{person}{Eric Sigler}, \bibinfo{person}{Mateusz Litwin}, \bibinfo{person}{Scott Gray}, \bibinfo{person}{Benjamin Chess}, \bibinfo{person}{Jack Clark}, \bibinfo{person}{Christopher Berner}, \bibinfo{person}{Sam McCandlish}, \bibinfo{person}{Alec Radford}, \bibinfo{person}{Ilya Sutskever},
  {and} \bibinfo{person}{Dario Amodei}.} \bibinfo{year}{2020}\natexlab{}.
\newblock \bibinfo{title}{Language Models are Few-Shot Learners}.
\newblock
\showeprint[arxiv]{2005.14165}~[cs.CL]
\urldef\tempurl%
\url{https://arxiv.org/abs/2005.14165}
\showURL{%
\tempurl}


\bibitem[Chapman et~al\mbox{.}(2001)]%
        {chapman2001simple}
\bibfield{author}{\bibinfo{person}{Wendy~W Chapman}, \bibinfo{person}{Will Bridewell}, \bibinfo{person}{Paul Hanbury}, \bibinfo{person}{Gregory~F Cooper}, {and} \bibinfo{person}{Bruce~G Buchanan}.} \bibinfo{year}{2001}\natexlab{}.
\newblock \showarticletitle{A simple algorithm for identifying negated findings and diseases in discharge summaries}.
\newblock \bibinfo{journal}{\emph{Journal of biomedical informatics}} \bibinfo{volume}{34}, \bibinfo{number}{5} (\bibinfo{year}{2001}), \bibinfo{pages}{301--310}.
\newblock


\bibitem[Chen et~al\mbox{.}(2024)]%
        {baai_bge_reranker_v2_m3}
\bibfield{author}{\bibinfo{person}{Jianlv Chen}, \bibinfo{person}{Shitao Xiao}, \bibinfo{person}{Peitian Zhang}, \bibinfo{person}{Kun Luo}, \bibinfo{person}{Defu Lian}, {and} \bibinfo{person}{Zheng Liu}.} \bibinfo{year}{2024}\natexlab{}.
\newblock \bibinfo{title}{BGE M3-Embedding: Multi-Lingual, Multi-Functionality, Multi-Granularity Text Embeddings Through Self-Knowledge Distillation}.
\newblock
\showeprint[arxiv]{2402.03216}~[cs.CL]
\urldef\tempurl%
\url{https://arxiv.org/abs/2402.03216}
\showURL{%
\tempurl}


\bibitem[Dubey et~al\mbox{.}(2024)]%
        {llama3}
\bibfield{author}{\bibinfo{person}{Abhimanyu Dubey}, \bibinfo{person}{Abhinav Jauhri}, \bibinfo{person}{Abhinav Pandey}, \bibinfo{person}{Abhishek Kadian}, \bibinfo{person}{Ahmad Al-Dahle}, \bibinfo{person}{Aiesha Letman}, \bibinfo{person}{Akhil Mathur}, \bibinfo{person}{Alan Schelten}, \bibinfo{person}{Amy Yang}, \bibinfo{person}{Angela Fan}, {et~al\mbox{.}}} \bibinfo{year}{2024}\natexlab{}.
\newblock \showarticletitle{The llama 3 herd of models}.
\newblock \bibinfo{journal}{\emph{arXiv preprint arXiv:2407.21783}} (\bibinfo{year}{2024}).
\newblock


\bibitem[Formal et~al\mbox{.}(2021)]%
        {formal2021splade}
\bibfield{author}{\bibinfo{person}{Thibault Formal}, \bibinfo{person}{Carlos Lassance}, \bibinfo{person}{Benjamin Piwowarski}, {and} \bibinfo{person}{St{\'e}phane Clinchant}.} \bibinfo{year}{2021}\natexlab{}.
\newblock \showarticletitle{SPLADE v2: Sparse lexical and expansion model for information retrieval}.
\newblock \bibinfo{journal}{\emph{arXiv preprint arXiv:2109.10086}} (\bibinfo{year}{2021}).
\newblock


\bibitem[Formal et~al\mbox{.}(2022)]%
        {formal2022distillation}
\bibfield{author}{\bibinfo{person}{Thibault Formal}, \bibinfo{person}{Carlos Lassance}, \bibinfo{person}{Benjamin Piwowarski}, {and} \bibinfo{person}{St\'{e}phane Clinchant}.} \bibinfo{year}{2022}\natexlab{}.
\newblock \showarticletitle{From Distillation to Hard Negative Sampling: Making Sparse Neural IR Models More Effective} \emph{(\bibinfo{series}{SIGIR '22})}. \bibinfo{publisher}{Association for Computing Machinery}, \bibinfo{address}{New York, NY, USA}, \bibinfo{pages}{2353–2359}.
\newblock
\showISBNx{9781450387323}
\href{https://doi.org/10.1145/3477495.3531857}{doi:\nolinkurl{10.1145/3477495.3531857}}


\bibitem[Gangi~Reddy et~al\mbox{.}(2024)]%
        {FIRST}
\bibfield{author}{\bibinfo{person}{Revanth Gangi~Reddy}, \bibinfo{person}{JaeHyeok Doo}, \bibinfo{person}{Yifei Xu}, \bibinfo{person}{Md~Arafat Sultan}, \bibinfo{person}{Deevya Swain}, \bibinfo{person}{Avirup Sil}, {and} \bibinfo{person}{Heng Ji}.} \bibinfo{year}{2024}\natexlab{}.
\newblock \showarticletitle{{FIRST}: Faster Improved Listwise Reranking with Single Token Decoding}. In \bibinfo{booktitle}{\emph{Proceedings of the 2024 Conference on Empirical Methods in Natural Language Processing}}, \bibfield{editor}{\bibinfo{person}{Yaser Al-Onaizan}, \bibinfo{person}{Mohit Bansal}, {and} \bibinfo{person}{Yun-Nung Chen}} (Eds.). \bibinfo{publisher}{Association for Computational Linguistics}, \bibinfo{address}{Miami, Florida, USA}, \bibinfo{pages}{8642--8652}.
\newblock
\href{https://doi.org/10.18653/v1/2024.emnlp-main.491}{doi:\nolinkurl{10.18653/v1/2024.emnlp-main.491}}


\bibitem[Gao and Callan(2022)]%
        {gao-callan-2022-unsupervised}
\bibfield{author}{\bibinfo{person}{Luyu Gao} {and} \bibinfo{person}{Jamie Callan}.} \bibinfo{year}{2022}\natexlab{}.
\newblock \showarticletitle{Unsupervised Corpus Aware Language Model Pre-training for Dense Passage Retrieval}. In \bibinfo{booktitle}{\emph{Proceedings of the 60th Annual Meeting of the Association for Computational Linguistics (Volume 1: Long Papers)}}, \bibfield{editor}{\bibinfo{person}{Smaranda Muresan}, \bibinfo{person}{Preslav Nakov}, {and} \bibinfo{person}{Aline Villavicencio}} (Eds.). \bibinfo{publisher}{Association for Computational Linguistics}, \bibinfo{address}{Dublin, Ireland}, \bibinfo{pages}{2843--2853}.
\newblock
\href{https://doi.org/10.18653/v1/2022.acl-long.203}{doi:\nolinkurl{10.18653/v1/2022.acl-long.203}}


\bibitem[Garc{\'\i}a-Ferrero et~al\mbox{.}(2023)]%
        {garcia2023not}
\bibfield{author}{\bibinfo{person}{Iker Garc{\'\i}a-Ferrero}, \bibinfo{person}{Bego{\~n}a Altuna}, \bibinfo{person}{Javier {\'A}lvez}, \bibinfo{person}{Itziar Gonzalez-Dios}, {and} \bibinfo{person}{German Rigau}.} \bibinfo{year}{2023}\natexlab{}.
\newblock \showarticletitle{This is not a dataset: A large negation benchmark to challenge large language models}.
\newblock \bibinfo{journal}{\emph{arXiv preprint arXiv:2310.15941}} (\bibinfo{year}{2023}).
\newblock


\bibitem[Goldin and Chapman(2003)]%
        {goldin2003learning}
\bibfield{author}{\bibinfo{person}{Ilya Goldin} {and} \bibinfo{person}{Wendy~W Chapman}.} \bibinfo{year}{2003}\natexlab{}.
\newblock \showarticletitle{Learning to detect negation with ‘not’in medical texts}. In \bibinfo{booktitle}{\emph{Proc workshop on text analysis and search for bioinformatics, ACM SIGIR}}.
\newblock


\bibitem[Harkema et~al\mbox{.}(2009)]%
        {harkema2009context}
\bibfield{author}{\bibinfo{person}{Henk Harkema}, \bibinfo{person}{John~N Dowling}, \bibinfo{person}{Tyler Thornblade}, {and} \bibinfo{person}{Wendy~W Chapman}.} \bibinfo{year}{2009}\natexlab{}.
\newblock \showarticletitle{ConText: an algorithm for determining negation, experiencer, and temporal status from clinical reports}.
\newblock \bibinfo{journal}{\emph{Journal of biomedical informatics}} \bibinfo{volume}{42}, \bibinfo{number}{5} (\bibinfo{year}{2009}), \bibinfo{pages}{839--851}.
\newblock


\bibitem[Hofst\"{a}tter et~al\mbox{.}(2021)]%
        {hofstatter2021efficiently}
\bibfield{author}{\bibinfo{person}{Sebastian Hofst\"{a}tter}, \bibinfo{person}{Sheng-Chieh Lin}, \bibinfo{person}{Jheng-Hong Yang}, \bibinfo{person}{Jimmy Lin}, {and} \bibinfo{person}{Allan Hanbury}.} \bibinfo{year}{2021}\natexlab{}.
\newblock \showarticletitle{Efficiently Teaching an Effective Dense Retriever with Balanced Topic Aware Sampling} \emph{(\bibinfo{series}{SIGIR '21})}. \bibinfo{publisher}{Association for Computing Machinery}, \bibinfo{address}{New York, NY, USA}, \bibinfo{pages}{113–122}.
\newblock
\showISBNx{9781450380379}
\href{https://doi.org/10.1145/3404835.3462891}{doi:\nolinkurl{10.1145/3404835.3462891}}


\bibitem[Hossain et~al\mbox{.}(2022)]%
        {hossain2022analysis}
\bibfield{author}{\bibinfo{person}{Md~Mosharaf Hossain}, \bibinfo{person}{Dhivya Chinnappa}, {and} \bibinfo{person}{Eduardo Blanco}.} \bibinfo{year}{2022}\natexlab{}.
\newblock \showarticletitle{An analysis of negation in natural language understanding corpora}.
\newblock \bibinfo{journal}{\emph{arXiv preprint arXiv:2203.08929}} (\bibinfo{year}{2022}).
\newblock


\bibitem[Ivison et~al\mbox{.}(2023)]%
        {tulu2}
\bibfield{author}{\bibinfo{person}{Hamish Ivison}, \bibinfo{person}{Yizhong Wang}, \bibinfo{person}{Valentina Pyatkin}, \bibinfo{person}{Nathan Lambert}, \bibinfo{person}{Matthew Peters}, \bibinfo{person}{Pradeep Dasigi}, \bibinfo{person}{Joel Jang}, \bibinfo{person}{David Wadden}, \bibinfo{person}{Noah~A Smith}, \bibinfo{person}{Iz Beltagy}, {et~al\mbox{.}}} \bibinfo{year}{2023}\natexlab{}.
\newblock \showarticletitle{Camels in a changing climate: Enhancing lm adaptation with tulu 2}.
\newblock \bibinfo{journal}{\emph{arXiv preprint arXiv:2311.10702}} (\bibinfo{year}{2023}).
\newblock


\bibitem[Jiang et~al\mbox{.}(2023)]%
        {mistral7b}
\bibfield{author}{\bibinfo{person}{Albert~Q. Jiang}, \bibinfo{person}{Alexandre Sablayrolles}, \bibinfo{person}{Arthur Mensch}, \bibinfo{person}{Chris Bamford}, \bibinfo{person}{Devendra~Singh Chaplot}, \bibinfo{person}{Guillaume Lample}, \bibinfo{person}{Marc'Aurelio Ranzato}, \bibinfo{person}{Gabriel Synnaeve}, {and} \bibinfo{person}{Nicolas Usunier}.} \bibinfo{year}{2023}\natexlab{}.
\newblock \showarticletitle{Mistral 7B}.
\newblock \bibinfo{journal}{\emph{arXiv preprint arXiv:2310.06825}} (\bibinfo{year}{2023}).
\newblock
\urldef\tempurl%
\url{https://arxiv.org/abs/2310.06825}
\showURL{%
\tempurl}


\bibitem[Karpukhin et~al\mbox{.}(2020)]%
        {dpr}
\bibfield{author}{\bibinfo{person}{Vladimir Karpukhin}, \bibinfo{person}{Barlas Oğuz}, \bibinfo{person}{Sewon Min}, \bibinfo{person}{Patrick Lewis}, \bibinfo{person}{Ledell Wu}, \bibinfo{person}{Sergey Edunov}, \bibinfo{person}{Danqi Chen}, {and} \bibinfo{person}{{Wen Tau} Yih}.} \bibinfo{year}{2020}\natexlab{}.
\newblock \showarticletitle{Dense passage retrieval for open-domain question answering}. In \bibinfo{booktitle}{\emph{EMNLP 2020 - 2020 Conference on Empirical Methods in Natural Language Processing, Proceedings of the Conference}} \emph{(\bibinfo{series}{EMNLP 2020 - 2020 Conference on Empirical Methods in Natural Language Processing, Proceedings of the Conference})}. \bibinfo{publisher}{Association for Computational Linguistics (ACL)}, \bibinfo{pages}{6769--6781}.
\newblock
\newblock
\shownote{Publisher Copyright: {\textcopyright} 2020 Association for Computational Linguistics; 2020 Conference on Empirical Methods in Natural Language Processing, EMNLP 2020 ; Conference date: 16-11-2020 Through 20-11-2020}.


\bibitem[Kenton and Toutanova(2019)]%
        {BERT}
\bibfield{author}{\bibinfo{person}{Jacob Devlin Ming-Wei~Chang Kenton} {and} \bibinfo{person}{Lee~Kristina Toutanova}.} \bibinfo{year}{2019}\natexlab{}.
\newblock \showarticletitle{Bert: Pre-training of deep bidirectional transformers for language understanding}. In \bibinfo{booktitle}{\emph{Proceedings of naacL-HLT}}, Vol.~\bibinfo{volume}{1}. Minneapolis, Minnesota.
\newblock


\bibitem[Khattab and Zaharia(2020)]%
        {colbertv1}
\bibfield{author}{\bibinfo{person}{Omar Khattab} {and} \bibinfo{person}{Matei Zaharia}.} \bibinfo{year}{2020}\natexlab{}.
\newblock \showarticletitle{ColBERT: Efficient and Effective Passage Search via Contextualized Late Interaction over BERT}. In \bibinfo{booktitle}{\emph{Proceedings of the 43rd International ACM SIGIR Conference on Research and Development in Information Retrieval}} (Virtual Event, China) \emph{(\bibinfo{series}{SIGIR '20})}. \bibinfo{publisher}{Association for Computing Machinery}, \bibinfo{address}{New York, NY, USA}, \bibinfo{pages}{39–48}.
\newblock
\showISBNx{9781450380164}
\urldef\tempurl%
\url{https://doi.org/10.1145/3397271.3401075}
\showURL{%
\tempurl}


\bibitem[Lacoste et~al\mbox{.}(2019)]%
        {lacoste2019quantifying}
\bibfield{author}{\bibinfo{person}{Alexandre Lacoste}, \bibinfo{person}{Alexandra Luccioni}, \bibinfo{person}{Victor Schmidt}, {and} \bibinfo{person}{Thomas Dandres}.} \bibinfo{year}{2019}\natexlab{}.
\newblock \showarticletitle{Quantifying the Carbon Emissions of Machine Learning}.
\newblock \bibinfo{journal}{\emph{arXiv preprint arXiv:1910.09700}} (\bibinfo{year}{2019}).
\newblock


\bibitem[Lassance and Clinchant(2022)]%
        {lassance2022efficiency}
\bibfield{author}{\bibinfo{person}{Carlos Lassance} {and} \bibinfo{person}{St\'{e}phane Clinchant}.} \bibinfo{year}{2022}\natexlab{}.
\newblock \showarticletitle{An Efficiency Study for SPLADE Models}. In \bibinfo{booktitle}{\emph{Proceedings of the 45th International ACM SIGIR Conference on Research and Development in Information Retrieval}} (Madrid, Spain) \emph{(\bibinfo{series}{SIGIR '22})}. \bibinfo{publisher}{Association for Computing Machinery}, \bibinfo{address}{New York, NY, USA}, \bibinfo{pages}{2220–2226}.
\newblock
\showISBNx{9781450387323}
\href{https://doi.org/10.1145/3477495.3531833}{doi:\nolinkurl{10.1145/3477495.3531833}}


\bibitem[Lassance et~al\mbox{.}(2024)]%
        {spladev3}
\bibfield{author}{\bibinfo{person}{Carlos Lassance}, \bibinfo{person}{Hervé Déjean}, \bibinfo{person}{Thibault Formal}, {and} \bibinfo{person}{Stéphane Clinchant}.} \bibinfo{year}{2024}\natexlab{}.
\newblock \bibinfo{title}{SPLADE-v3: New baselines for SPLADE}.
\newblock
\showeprint[arxiv]{2403.06789}~[cs.IR]
\urldef\tempurl%
\url{https://arxiv.org/abs/2403.06789}
\showURL{%
\tempurl}


\bibitem[Li et~al\mbox{.}(2023)]%
        {qwen_gte}
\bibfield{author}{\bibinfo{person}{Zehan Li}, \bibinfo{person}{Xin Zhang}, \bibinfo{person}{Yanzhao Zhang}, \bibinfo{person}{Dingkun Long}, \bibinfo{person}{Pengjun Xie}, {and} \bibinfo{person}{Meishan Zhang}.} \bibinfo{year}{2023}\natexlab{}.
\newblock \showarticletitle{Towards general text embeddings with multi-stage contrastive learning}.
\newblock \bibinfo{journal}{\emph{arXiv preprint arXiv:2308.03281}} (\bibinfo{year}{2023}).
\newblock


\bibitem[Lin et~al\mbox{.}(2023)]%
        {dragon}
\bibfield{author}{\bibinfo{person}{Sheng-Chieh Lin}, \bibinfo{person}{Akari Asai}, \bibinfo{person}{Minghan Li}, \bibinfo{person}{Barlas Oguz}, \bibinfo{person}{Jimmy Lin}, \bibinfo{person}{Yashar Mehdad}, \bibinfo{person}{Wen-tau Yih}, {and} \bibinfo{person}{Xilun Chen}.} \bibinfo{year}{2023}\natexlab{}.
\newblock \showarticletitle{How to Train Your Dragon: Diverse Augmentation Towards Generalizable Dense Retrieval}. In \bibinfo{booktitle}{\emph{Findings of the Association for Computational Linguistics: EMNLP 2023}}, \bibfield{editor}{\bibinfo{person}{Houda Bouamor}, \bibinfo{person}{Juan Pino}, {and} \bibinfo{person}{Kalika Bali}} (Eds.). \bibinfo{publisher}{Association for Computational Linguistics}, \bibinfo{address}{Singapore}, \bibinfo{pages}{6385--6400}.
\newblock
\href{https://doi.org/10.18653/v1/2023.findings-emnlp.423}{doi:\nolinkurl{10.18653/v1/2023.findings-emnlp.423}}


\bibitem[Ma et~al\mbox{.}(2024)]%
        {repllama_rankllama}
\bibfield{author}{\bibinfo{person}{Xueguang Ma}, \bibinfo{person}{Liang Wang}, \bibinfo{person}{Nan Yang}, \bibinfo{person}{Furu Wei}, {and} \bibinfo{person}{Jimmy Lin}.} \bibinfo{year}{2024}\natexlab{}.
\newblock \showarticletitle{Fine-Tuning LLaMA for Multi-Stage Text Retrieval}. In \bibinfo{booktitle}{\emph{Proceedings of the 47th International ACM SIGIR Conference on Research and Development in Information Retrieval}} (Washington DC, USA) \emph{(\bibinfo{series}{SIGIR '24})}. \bibinfo{publisher}{Association for Computing Machinery}, \bibinfo{address}{New York, NY, USA}, \bibinfo{pages}{2421–2425}.
\newblock
\showISBNx{9798400704314}
\href{https://doi.org/10.1145/3626772.3657951}{doi:\nolinkurl{10.1145/3626772.3657951}}


\bibitem[Ma et~al\mbox{.}(2023)]%
        {ma2023zeroshotlistwisedocumentreranking}
\bibfield{author}{\bibinfo{person}{Xueguang Ma}, \bibinfo{person}{Xinyu Zhang}, \bibinfo{person}{Ronak Pradeep}, {and} \bibinfo{person}{Jimmy Lin}.} \bibinfo{year}{2023}\natexlab{}.
\newblock \bibinfo{title}{Zero-Shot Listwise Document Reranking with a Large Language Model}.
\newblock
\showeprint[arxiv]{2305.02156}~[cs.IR]
\urldef\tempurl%
\url{https://arxiv.org/abs/2305.02156}
\showURL{%
\tempurl}


\bibitem[Malaviya et~al\mbox{.}(2023)]%
        {malaviya2023quest}
\bibfield{author}{\bibinfo{person}{Chaitanya Malaviya}, \bibinfo{person}{Peter Shaw}, \bibinfo{person}{Ming-Wei Chang}, \bibinfo{person}{Kenton Lee}, {and} \bibinfo{person}{Kristina Toutanova}.} \bibinfo{year}{2023}\natexlab{}.
\newblock \showarticletitle{{QUEST}: A Retrieval Dataset of Entity-Seeking Queries with Implicit Set Operations}. In \bibinfo{booktitle}{\emph{Proceedings of the 61st Annual Meeting of the Association for Computational Linguistics (Volume 1: Long Papers)}}, \bibfield{editor}{\bibinfo{person}{Anna Rogers}, \bibinfo{person}{Jordan Boyd-Graber}, {and} \bibinfo{person}{Naoaki Okazaki}} (Eds.). \bibinfo{publisher}{Association for Computational Linguistics}, \bibinfo{address}{Toronto, Canada}, \bibinfo{pages}{14032--14047}.
\newblock
\href{https://doi.org/10.18653/v1/2023.acl-long.784}{doi:\nolinkurl{10.18653/v1/2023.acl-long.784}}


\bibitem[Morante and Blanco(2021)]%
        {morante2021recent}
\bibfield{author}{\bibinfo{person}{Roser Morante} {and} \bibinfo{person}{Eduardo Blanco}.} \bibinfo{year}{2021}\natexlab{}.
\newblock \showarticletitle{Recent advances in processing negation}.
\newblock \bibinfo{journal}{\emph{Natural Language Engineering}} \bibinfo{volume}{27}, \bibinfo{number}{2} (\bibinfo{year}{2021}), \bibinfo{pages}{121--130}.
\newblock


\bibitem[Muennighoff et~al\mbox{.}(2024)]%
        {gritlm}
\bibfield{author}{\bibinfo{person}{Niklas Muennighoff}, \bibinfo{person}{Hongjin Su}, \bibinfo{person}{Liang Wang}, \bibinfo{person}{Nan Yang}, \bibinfo{person}{Furu Wei}, \bibinfo{person}{Tao Yu}, \bibinfo{person}{Amanpreet Singh}, {and} \bibinfo{person}{Douwe Kiela}.} \bibinfo{year}{2024}\natexlab{}.
\newblock \bibinfo{title}{Generative Representational Instruction Tuning}.
\newblock
\showeprint[arxiv]{2402.09906}~[cs.CL]


\bibitem[Nogueira et~al\mbox{.}(2020)]%
        {nogueira-etal-2020-document}
\bibfield{author}{\bibinfo{person}{Rodrigo Nogueira}, \bibinfo{person}{Zhiying Jiang}, \bibinfo{person}{Ronak Pradeep}, {and} \bibinfo{person}{Jimmy Lin}.} \bibinfo{year}{2020}\natexlab{}.
\newblock \showarticletitle{Document Ranking with a Pretrained Sequence-to-Sequence Model}. In \bibinfo{booktitle}{\emph{Findings of the Association for Computational Linguistics: EMNLP 2020}}, \bibfield{editor}{\bibinfo{person}{Trevor Cohn}, \bibinfo{person}{Yulan He}, {and} \bibinfo{person}{Yang Liu}} (Eds.). \bibinfo{publisher}{Association for Computational Linguistics}, \bibinfo{address}{Online}, \bibinfo{pages}{708--718}.
\newblock
\href{https://doi.org/10.18653/v1/2020.findings-emnlp.63}{doi:\nolinkurl{10.18653/v1/2020.findings-emnlp.63}}


\bibitem[OpenAI(2023)]%
        {openai2024gpt4technicalreport}
\bibfield{author}{\bibinfo{person}{OpenAI}.} \bibinfo{year}{2023}\natexlab{}.
\newblock \showarticletitle{{GPT-4} Technical Report}.
\newblock \bibinfo{journal}{\emph{CoRR}}  \bibinfo{volume}{abs/2303.08774} (\bibinfo{year}{2023}).
\newblock
\href{https://doi.org/10.48550/ARXIV.2303.08774}{doi:\nolinkurl{10.48550/ARXIV.2303.08774}}
\showeprint[arXiv]{2303.08774}


\bibitem[Peng et~al\mbox{.}(2018)]%
        {peng2018negbio}
\bibfield{author}{\bibinfo{person}{Yifan Peng}, \bibinfo{person}{Xiaosong Wang}, \bibinfo{person}{Le Lu}, \bibinfo{person}{Mohammadhadi Bagheri}, \bibinfo{person}{Ronald Summers}, {and} \bibinfo{person}{Zhiyong Lu}.} \bibinfo{year}{2018}\natexlab{}.
\newblock \showarticletitle{NegBio: a high-performance tool for negation and uncertainty detection in radiology reports}.
\newblock \bibinfo{journal}{\emph{AMIA Summits on Translational Science Proceedings}}  \bibinfo{volume}{2018} (\bibinfo{year}{2018}), \bibinfo{pages}{188}.
\newblock


\bibitem[Pradeep et~al\mbox{.}(2023)]%
        {rankLLM}
\bibfield{author}{\bibinfo{person}{Ronak Pradeep}, \bibinfo{person}{Sahel Sharifymoghaddam}, {and} \bibinfo{person}{Jimmy Lin}.} \bibinfo{year}{2023}\natexlab{}.
\newblock \showarticletitle{Rankvicuna: Zero-shot listwise document reranking with open-source large language models}.
\newblock \bibinfo{journal}{\emph{arXiv preprint arXiv:2309.15088}} (\bibinfo{year}{2023}).
\newblock


\bibitem[Qin et~al\mbox{.}(2024)]%
        {qin2023large}
\bibfield{author}{\bibinfo{person}{Zhen Qin}, \bibinfo{person}{Rolf Jagerman}, \bibinfo{person}{Kai Hui}, \bibinfo{person}{Honglei Zhuang}, \bibinfo{person}{Junru Wu}, \bibinfo{person}{Le Yan}, \bibinfo{person}{Jiaming Shen}, \bibinfo{person}{Tianqi Liu}, \bibinfo{person}{Jialu Liu}, \bibinfo{person}{Donald Metzler}, \bibinfo{person}{Xuanhui Wang}, {and} \bibinfo{person}{Michael Bendersky}.} \bibinfo{year}{2024}\natexlab{}.
\newblock \showarticletitle{Large Language Models are Effective Text Rankers with Pairwise Ranking Prompting}. In \bibinfo{booktitle}{\emph{Findings of the Association for Computational Linguistics: NAACL 2024}}, \bibfield{editor}{\bibinfo{person}{Kevin Duh}, \bibinfo{person}{Helena Gomez}, {and} \bibinfo{person}{Steven Bethard}} (Eds.). \bibinfo{publisher}{Association for Computational Linguistics}, \bibinfo{address}{Mexico City, Mexico}, \bibinfo{pages}{1504--1518}.
\newblock
\href{https://doi.org/10.18653/v1/2024.findings-naacl.97}{doi:\nolinkurl{10.18653/v1/2024.findings-naacl.97}}


\bibitem[Qu et~al\mbox{.}(2021)]%
        {qu-etal-2021-rocketqa}
\bibfield{author}{\bibinfo{person}{Yingqi Qu}, \bibinfo{person}{Yuchen Ding}, \bibinfo{person}{Jing Liu}, \bibinfo{person}{Kai Liu}, \bibinfo{person}{Ruiyang Ren}, \bibinfo{person}{Wayne~Xin Zhao}, \bibinfo{person}{Daxiang Dong}, \bibinfo{person}{Hua Wu}, {and} \bibinfo{person}{Haifeng Wang}.} \bibinfo{year}{2021}\natexlab{}.
\newblock \showarticletitle{{R}ocket{QA}: An Optimized Training Approach to Dense Passage Retrieval for Open-Domain Question Answering}. In \bibinfo{booktitle}{\emph{Proceedings of the 2021 Conference of the North American Chapter of the Association for Computational Linguistics: Human Language Technologies}}, \bibfield{editor}{\bibinfo{person}{Kristina Toutanova}, \bibinfo{person}{Anna Rumshisky}, \bibinfo{person}{Luke Zettlemoyer}, \bibinfo{person}{Dilek Hakkani-Tur}, \bibinfo{person}{Iz~Beltagy}, \bibinfo{person}{Steven Bethard}, \bibinfo{person}{Ryan Cotterell}, \bibinfo{person}{Tanmoy Chakraborty}, {and} \bibinfo{person}{Yichao Zhou}} (Eds.). \bibinfo{publisher}{Association for Computational Linguistics}, \bibinfo{address}{Online}, \bibinfo{pages}{5835--5847}.
\newblock
\href{https://doi.org/10.18653/v1/2021.naacl-main.466}{doi:\nolinkurl{10.18653/v1/2021.naacl-main.466}}


\bibitem[Raffel et~al\mbox{.}(2020)]%
        {T5}
\bibfield{author}{\bibinfo{person}{Colin Raffel}, \bibinfo{person}{Noam Shazeer}, \bibinfo{person}{Adam Roberts}, \bibinfo{person}{Katherine Lee}, \bibinfo{person}{Sharan Narang}, \bibinfo{person}{Michael Matena}, \bibinfo{person}{Yanqi Zhou}, \bibinfo{person}{Wei Li}, {and} \bibinfo{person}{Peter~J Liu}.} \bibinfo{year}{2020}\natexlab{}.
\newblock \showarticletitle{Exploring the limits of transfer learning with a unified text-to-text transformer}.
\newblock \bibinfo{journal}{\emph{Journal of machine learning research}} \bibinfo{volume}{21}, \bibinfo{number}{140} (\bibinfo{year}{2020}), \bibinfo{pages}{1--67}.
\newblock


\bibitem[Ravichander et~al\mbox{.}(2022)]%
        {condaqa}
\bibfield{author}{\bibinfo{person}{Abhilasha Ravichander}, \bibinfo{person}{Matt Gardner}, {and} \bibinfo{person}{Ana Marasovi{\'c}}.} \bibinfo{year}{2022}\natexlab{}.
\newblock \showarticletitle{CONDAQA: A Contrastive Reading Comprehension Dataset for Reasoning about Negation}. In \bibinfo{booktitle}{\emph{Proceedings of the 2022 Conference on Empirical Methods in Natural Language Processing}}. \bibinfo{pages}{8729--8755}.
\newblock


\bibitem[Reimers(2019)]%
        {sentencetransformers}
\bibfield{author}{\bibinfo{person}{N Reimers}.} \bibinfo{year}{2019}\natexlab{}.
\newblock \showarticletitle{Sentence-BERT: Sentence Embeddings using Siamese BERT-Networks}.
\newblock \bibinfo{journal}{\emph{arXiv preprint arXiv:1908.10084}} (\bibinfo{year}{2019}).
\newblock


\bibitem[Reimers and Gurevych(2019)]%
        {reimers-gurevych-2019-sentence}
\bibfield{author}{\bibinfo{person}{Nils Reimers} {and} \bibinfo{person}{Iryna Gurevych}.} \bibinfo{year}{2019}\natexlab{}.
\newblock \showarticletitle{Sentence-{BERT}: Sentence Embeddings using {S}iamese {BERT}-Networks}. In \bibinfo{booktitle}{\emph{Proceedings of the 2019 Conference on Empirical Methods in Natural Language Processing and the 9th International Joint Conference on Natural Language Processing (EMNLP-IJCNLP)}}, \bibfield{editor}{\bibinfo{person}{Kentaro Inui}, \bibinfo{person}{Jing Jiang}, \bibinfo{person}{Vincent Ng}, {and} \bibinfo{person}{Xiaojun Wan}} (Eds.). \bibinfo{publisher}{Association for Computational Linguistics}, \bibinfo{address}{Hong Kong, China}, \bibinfo{pages}{3982--3992}.
\newblock
\href{https://doi.org/10.18653/v1/D19-1410}{doi:\nolinkurl{10.18653/v1/D19-1410}}


\bibitem[Robertson et~al\mbox{.}(2009)]%
        {bm25}
\bibfield{author}{\bibinfo{person}{Stephen Robertson}, \bibinfo{person}{Hugo Zaragoza}, {et~al\mbox{.}}} \bibinfo{year}{2009}\natexlab{}.
\newblock \showarticletitle{The probabilistic relevance framework: BM25 and beyond}.
\newblock \bibinfo{journal}{\emph{Foundations and Trends{\textregistered} in Information Retrieval}} \bibinfo{volume}{3}, \bibinfo{number}{4} (\bibinfo{year}{2009}), \bibinfo{pages}{333--389}.
\newblock


\bibitem[Santhanam et~al\mbox{.}(2022)]%
        {santhanam2021colbertv2}
\bibfield{author}{\bibinfo{person}{Keshav Santhanam}, \bibinfo{person}{Omar Khattab}, \bibinfo{person}{Jon Saad-Falcon}, \bibinfo{person}{Christopher Potts}, {and} \bibinfo{person}{Matei Zaharia}.} \bibinfo{year}{2022}\natexlab{}.
\newblock \showarticletitle{{C}ol{BERT}v2: Effective and Efficient Retrieval via Lightweight Late Interaction}. In \bibinfo{booktitle}{\emph{Proceedings of the 2022 Conference of the North American Chapter of the Association for Computational Linguistics: Human Language Technologies}}, \bibfield{editor}{\bibinfo{person}{Marine Carpuat}, \bibinfo{person}{Marie-Catherine de~Marneffe}, {and} \bibinfo{person}{Ivan~Vladimir Meza~Ruiz}} (Eds.). \bibinfo{publisher}{Association for Computational Linguistics}, \bibinfo{address}{Seattle, United States}, \bibinfo{pages}{3715--3734}.
\newblock
\href{https://doi.org/10.18653/v1/2022.naacl-main.272}{doi:\nolinkurl{10.18653/v1/2022.naacl-main.272}}


\bibitem[Sineva et~al\mbox{.}(2021)]%
        {sineva2021negation}
\bibfield{author}{\bibinfo{person}{Elizaveta Sineva}, \bibinfo{person}{Stefan Gr{\"u}newald}, \bibinfo{person}{Annemarie Friedrich}, {and} \bibinfo{person}{Jonas Kuhn}.} \bibinfo{year}{2021}\natexlab{}.
\newblock \showarticletitle{Negation-instance based evaluation of end-to-end negation resolution}.
\newblock \bibinfo{journal}{\emph{arXiv preprint arXiv:2109.10013}} (\bibinfo{year}{2021}).
\newblock


\bibitem[Sun et~al\mbox{.}(2023)]%
        {rankGPT}
\bibfield{author}{\bibinfo{person}{Weiwei Sun}, \bibinfo{person}{Lingyong Yan}, \bibinfo{person}{Xinyu Ma}, \bibinfo{person}{Shuaiqiang Wang}, \bibinfo{person}{Pengjie Ren}, \bibinfo{person}{Zhumin Chen}, \bibinfo{person}{Dawei Yin}, {and} \bibinfo{person}{Zhaochun Ren}.} \bibinfo{year}{2023}\natexlab{}.
\newblock \showarticletitle{Is ChatGPT Good at Search? Investigating Large Language Models as Re-Ranking Agents}. In \bibinfo{booktitle}{\emph{Proceedings of the 2023 Conference on Empirical Methods in Natural Language Processing}}. \bibinfo{pages}{14918--14937}.
\newblock


\bibitem[Taylor and Harabagiu(2018)]%
        {taylor2018role}
\bibfield{author}{\bibinfo{person}{Stuart~J Taylor} {and} \bibinfo{person}{Sanda~M Harabagiu}.} \bibinfo{year}{2018}\natexlab{}.
\newblock \showarticletitle{The role of a deep-learning method for negation detection in patient cohort identification from electroencephalography reports}. In \bibinfo{booktitle}{\emph{AMIA Annual Symposium Proceedings}}, Vol.~\bibinfo{volume}{2018}. \bibinfo{pages}{1018}.
\newblock


\bibitem[Uzuner et~al\mbox{.}(2011)]%
        {uzuner20112010}
\bibfield{author}{\bibinfo{person}{{\"O}zlem Uzuner}, \bibinfo{person}{Brett~R South}, \bibinfo{person}{Shuying Shen}, {and} \bibinfo{person}{Scott~L DuVall}.} \bibinfo{year}{2011}\natexlab{}.
\newblock \showarticletitle{2010 i2b2/VA challenge on concepts, assertions, and relations in clinical text}.
\newblock \bibinfo{journal}{\emph{Journal of the American Medical Informatics Association}} \bibinfo{volume}{18}, \bibinfo{number}{5} (\bibinfo{year}{2011}), \bibinfo{pages}{552--556}.
\newblock


\bibitem[Wang et~al\mbox{.}(2024)]%
        {e5}
\bibfield{author}{\bibinfo{person}{Liang Wang}, \bibinfo{person}{Nan Yang}, \bibinfo{person}{Xiaolong Huang}, \bibinfo{person}{Linjun Yang}, \bibinfo{person}{Rangan Majumder}, {and} \bibinfo{person}{Furu Wei}.} \bibinfo{year}{2024}\natexlab{}.
\newblock \showarticletitle{Multilingual e5 text embeddings: A technical report}.
\newblock \bibinfo{journal}{\emph{arXiv preprint arXiv:2402.05672}} (\bibinfo{year}{2024}).
\newblock


\bibitem[Weller et~al\mbox{.}(2024a)]%
        {promptriever}
\bibfield{author}{\bibinfo{person}{Orion Weller}, \bibinfo{person}{Benjamin~Van Durme}, \bibinfo{person}{Dawn Lawrie}, \bibinfo{person}{Ashwin Paranjape}, \bibinfo{person}{Yuhao Zhang}, {and} \bibinfo{person}{Jack Hessel}.} \bibinfo{year}{2024}\natexlab{a}.
\newblock \bibinfo{title}{Promptriever: Instruction-Trained Retrievers Can Be Prompted Like Language Models}.
\newblock
\showeprint[arxiv]{2409.11136}~[cs.IR]
\urldef\tempurl%
\url{https://arxiv.org/abs/2409.11136}
\showURL{%
\tempurl}


\bibitem[Weller et~al\mbox{.}(2024b)]%
        {weller2024nevirnegationneuralinformation}
\bibfield{author}{\bibinfo{person}{Orion Weller}, \bibinfo{person}{Dawn Lawrie}, {and} \bibinfo{person}{Benjamin Van~Durme}.} \bibinfo{year}{2024}\natexlab{b}.
\newblock \showarticletitle{NevIR: Negation in Neural Information Retrieval}. In \bibinfo{booktitle}{\emph{Proceedings of the 18th Conference of the European Chapter of the Association for Computational Linguistics (Volume 1: Long Papers)}}. \bibinfo{pages}{2274--2287}.
\newblock


\bibitem[Yang et~al\mbox{.}(2024)]%
        {yang2024qwen2technicalreport}
\bibfield{author}{\bibinfo{person}{An Yang}, \bibinfo{person}{Baosong Yang}, \bibinfo{person}{Binyuan Hui}, \bibinfo{person}{Bo Zheng}, \bibinfo{person}{Bowen Yu}, \bibinfo{person}{Chang Zhou}, \bibinfo{person}{Chengpeng Li}, \bibinfo{person}{Chengyuan Li}, \bibinfo{person}{Dayiheng Liu}, \bibinfo{person}{Fei Huang}, \bibinfo{person}{Guanting Dong}, \bibinfo{person}{Haoran Wei}, \bibinfo{person}{Huan Lin}, \bibinfo{person}{Jialong Tang}, \bibinfo{person}{Jialin Wang}, \bibinfo{person}{Jian Yang}, \bibinfo{person}{Jianhong Tu}, \bibinfo{person}{Jianwei Zhang}, \bibinfo{person}{Jianxin Ma}, \bibinfo{person}{Jianxin Yang}, \bibinfo{person}{Jin Xu}, \bibinfo{person}{Jingren Zhou}, \bibinfo{person}{Jinze Bai}, \bibinfo{person}{Jinzheng He}, \bibinfo{person}{Junyang Lin}, \bibinfo{person}{Kai Dang}, \bibinfo{person}{Keming Lu}, \bibinfo{person}{Keqin Chen}, \bibinfo{person}{Kexin Yang}, \bibinfo{person}{Mei Li}, \bibinfo{person}{Mingfeng Xue}, \bibinfo{person}{Na Ni}, \bibinfo{person}{Pei Zhang},
  \bibinfo{person}{Peng Wang}, \bibinfo{person}{Ru Peng}, \bibinfo{person}{Rui Men}, \bibinfo{person}{Ruize Gao}, \bibinfo{person}{Runji Lin}, \bibinfo{person}{Shijie Wang}, \bibinfo{person}{Shuai Bai}, \bibinfo{person}{Sinan Tan}, \bibinfo{person}{Tianhang Zhu}, \bibinfo{person}{Tianhao Li}, \bibinfo{person}{Tianyu Liu}, \bibinfo{person}{Wenbin Ge}, \bibinfo{person}{Xiaodong Deng}, \bibinfo{person}{Xiaohuan Zhou}, \bibinfo{person}{Xingzhang Ren}, \bibinfo{person}{Xinyu Zhang}, \bibinfo{person}{Xipin Wei}, \bibinfo{person}{Xuancheng Ren}, \bibinfo{person}{Xuejing Liu}, \bibinfo{person}{Yang Fan}, \bibinfo{person}{Yang Yao}, \bibinfo{person}{Yichang Zhang}, \bibinfo{person}{Yu Wan}, \bibinfo{person}{Yunfei Chu}, \bibinfo{person}{Yuqiong Liu}, \bibinfo{person}{Zeyu Cui}, \bibinfo{person}{Zhenru Zhang}, \bibinfo{person}{Zhifang Guo}, {and} \bibinfo{person}{Zhihao Fan}.} \bibinfo{year}{2024}\natexlab{}.
\newblock \bibinfo{title}{Qwen2 Technical Report}.
\newblock
\showeprint[arxiv]{2407.10671}~[cs.CL]
\urldef\tempurl%
\url{https://arxiv.org/abs/2407.10671}
\showURL{%
\tempurl}


\bibitem[Zhang et~al\mbox{.}(2024)]%
        {zhang2024excluir}
\bibfield{author}{\bibinfo{person}{Wenhao Zhang}, \bibinfo{person}{Mengqi Zhang}, \bibinfo{person}{Shiguang Wu}, \bibinfo{person}{Jiahuan Pei}, \bibinfo{person}{Zhaochun Ren}, \bibinfo{person}{Maarten de Rijke}, \bibinfo{person}{Zhumin Chen}, {and} \bibinfo{person}{Pengjie Ren}.} \bibinfo{year}{2024}\natexlab{}.
\newblock \showarticletitle{ExcluIR: Exclusionary Neural Information Retrieval}.
\newblock \bibinfo{journal}{\emph{arXiv preprint arXiv:2404.17288}} (\bibinfo{year}{2024}).
\newblock


\bibitem[Zhu et~al\mbox{.}({[n.\,d.]})]%
        {zhularge}
\bibfield{author}{\bibinfo{person}{Yutao Zhu}, \bibinfo{person}{Huaying Yuan}, \bibinfo{person}{Shuting Wang}, \bibinfo{person}{Jiongnan Liu}, \bibinfo{person}{Wenhan Liu}, \bibinfo{person}{Chenlong Deng}, \bibinfo{person}{Haonan Chen}, \bibinfo{person}{Zhicheng Dou}, {and} \bibinfo{person}{Ji-Rong Wen}.} \bibinfo{year}{[n.\,d.]}\natexlab{}.
\newblock \showarticletitle{Large Language Models for Information Retrieval: A Survey}.
\newblock  (\bibinfo{year}{[n.\,d.]}).
\newblock


\end{thebibliography}
